\newcommand{\lsim }{{\lower0.8ex\hbox{$\buildrel <\over\sim$}}}
\newcommand{\gsim }{{\lower0.8ex\hbox{$\buildrel >\over\sim$}}}
\newcommand{\up}{\ifmmode{u^{\prime}} \else $u^{\prime}$\fi}
\newcommand{\gp}{\ifmmode{g^{\prime}} \else $g^{\prime}$\fi}
\newcommand{\rp}{\ifmmode{r^{\prime}} \else $r^{\prime}$\fi}
\newcommand{\ip}{\ifmmode{i^{\prime}} \else $i^{\prime}$\fi}
\newcommand{\zp}{\ifmmode{z^{\prime}} \else $z^{\prime}$\fi}
\newcommand{\us}{\ifmmode{u^{\ast}} \else $u^{\ast}$\fi}
\newcommand{\gs}{\ifmmode{g^{\ast}} \else $g^{\ast}$\fi}
\newcommand{\rs}{\ifmmode{r^{\ast}} \else $r^{\ast}$\fi}
\newcommand{\is}{\ifmmode{i^{\ast}} \else $i^{\ast}$\fi}
\newcommand{\zs}{\ifmmode{z^{\ast}} \else $z^{\ast}$\fi}
\newcommand{\gmr}{\ifmmode{(g^{\prime}-r^{\prime})} \else $(g^{\prime}-r^{\prime})$\fi}
\newcommand{\rmi}{\ifmmode{(r^{\prime}-i^{\prime})} \else $(r^{\prime}-i^{\prime})$\fi}
\newcommand{\gmrs}{\ifmmode{(g^{\ast}-r^{\ast})} \else $(g^{\ast}-r^{\ast})$\fi}
\newcommand{\rmis}{\ifmmode{(r^{\ast}-i^{\ast})} \else $(r^{\ast}-i^{\ast})$\fi}
\def\Chandra     {{\em Chandra}}
\newcommand{\fcgs}{\ifmmode {\rm erg~cm}^{-2}~{\rm s}^{-1}\else 
erg~cm$^{-2}$~s$^{-1}$\fi}
\newcommand{\lcgs}{\ifmmode {\rm erg~s}^{-1}\else erg~s$^{-1}$\fi}
\newcommand{\flamcgs}{\ifmmode {\rm erg~cm}^{-2}~{\rm s}^{-1}~Hz^{-1}\else 
erg~cm$^{-2}$~s$^{-1}~$\AA$^{-1}$\fi}
\newcommand{\fnucgs}{\ifmmode {\rm erg~cm}^{-2}~{\rm s}^{-1}~{\rm
Hz}^{-1}\else  erg~cm$^{-2}$~s$^{-1}$~Hz$^{-1}$\fi}
\newcommand{\lnucgs}{\ifmmode {\rm erg~s}^{-1}~{\rm Hz}^{-1}\else 
erg~s$^{-1}$~Hz$^{-1}$\fi}
\newcommand{\kms}{\ifmmode~{\rm km~s}^{-1}\else ~km~s$^{-1}~$\fi}
\newcommand{\mone}{\ifmmode ^{-1}\else$^{-1}$\fi}
\newcommand{\mtwo}{\ifmmode ^{-2}\else$^{-2}$\fi}
\newcommand{\degs}{\ifmmode ^{\circ}\else$^{\circ}$\fi}
\newcommand{\mv}{\ifmmode {m_{V}}\else${m_{V}}$\fi}
\newcommand{\Mv}{\ifmmode {M_{V}}\else${M_{V}}$\fi}
\newcommand{\msun}{\ifmmode {M_{\odot}}\else${M_{\odot}}$\fi}
\newcommand{\rsun}{\ifmmode {R_{\odot}}\else${R_{\odot}}$\fi}
\newcommand{\lsun}{\ifmmode {L_{\odot}}\else${L_{\odot}}$\fi}
\newcommand{\lapprox }{{\lower0.8ex\hbox{$\buildrel <\over\sim$}}}
\newcommand{\gapprox }{{\lower0.8ex\hbox{$\buildrel >\over\sim$}}}

\newcommand{\cmsq}{\ifmmode{\rm ~cm^{-2}} \else cm$^{-2}$\fi}
\newcommand{\ebmv}{\ifmmode{\rm E}_{B-V} \else E$_{B-V}$\fi}
\newcommand{\nh}{\ifmmode{\rm N_{H}} \else N$_{H}$\fi}
\newcommand{\nhgal}{\ifmmode{ N_{H}^{Gal}} \else N$_{H}^{Gal}$\fi}
\newcommand{\nhintr}{\ifmmode{ N_{H}^{intr}} \else N$_{H}^{intr}$\fi}
\newcommand{\nhtot}{\ifmmode{ N_{H}^{tot}} \else N$_{H}^{tot}$\fi}
\newcommand{\meangamma}{\ifmmode{\langle\Gamma\rangle} \else 
$\langle\Gamma\rangle$\fi}
\newcommand{\fx}{\ifmmode f_X \else $~f_X$\fi}
\newcommand{\fxfo}{\ifmmode \frac{f_X}{f_{opt}} \else
 $\frac{f_X}{f_{opt}}$\fi}
\newcommand{\fxfr}{\ifmmode \frac{f_X}{f_{r}} \else
 $\frac{f_X}{f_r}$\fi}
\newcommand{\logfx}{\ifmmode{\rm log}~f_X \else log$~f_X$\fi}
\newcommand{\logfxfo}{\ifmmode{\rm log}\,(\frac{f_X}{f_{opt}}) \else
 ${\rm log}\,(\frac{f_X}{f_{opt}})$ \fi}
\newcommand{\logxr}{\ifmmode{\rm log}\,\frac{f_X}{f_{r}} \else
 ${\rm log}\,\frac{f_X}{f_{r}}$ \fi}
\newcommand{\lopt}{\ifmmode L_{opt} \else $~L_{opt}$\fi}
\newcommand{\loglopt}{\ifmmode{\rm log}~L_{opt} \else log$~L_{opt}$\fi}
\newcommand{\lx}{\ifmmode L_X \else $~L_X$\fi}
\newcommand{\loglx}{\ifmmode{\rm log}~L_X \else log$~L_X$\fi}
\newcommand{\aox}{\ifmmode{\alpha_{ox}} \else $\alpha_{ox}$\fi} 
\newcommand{\snr}{\ifmmode{\frac{S}{N}} \else $\frac{S}{N}$\fi} 

\documentclass{emulateapj}

\slugcomment{To appear in The Astrophysical Journal}

\shorttitle{X-ray survey of AGN with Chandra}
\shortauthors{Silverman et al.}

\usepackage{lscape}
\begin{document}


\title{Hard X-ray emitting Active Galactic Nuclei selected by the
$Chandra$ Multi-wavelength Project}


\author{J. D. Silverman\altaffilmark{1,2,3}, P. J. Green\altaffilmark{4},
W. A. Barkhouse, D.-W. Kim, T. L. Aldcroft, R. A. Cameron, B.
J. Wilkes, A. Mossman, H. Ghosh, H. Tananbaum}
\affil{Harvard-Smithsonian Center for Astrophysics, 60 Garden Street ,
Cambridge, MA 02138}

\author{M. G. Smith, R. C. Smith} 

\affil{Cerro Tololo Inter-American Observatory, National Optical
Astronomical Observatory, Casilla 603, La Serena, Chile}

\author{P. S. Smith}
\affil{Steward Observatory, The University of Arizona, Tucson, AZ 85
721}

\author{C. Foltz}
\affil{National Science Foundation, 4201 Wilson Blvd., Arlington, VA, 22230}

\author{D. Wik}
\affil{Astronomy Department, University of Virginia, P.O. Box 3818, Charlottesville, VA, 22903-0818}

\and

\author{B. T. Jannuzi} \affil{National Optical
Astronomical Observatory, P.O. Box 26732, Tucson, AZ, 85726-6732}


\altaffiltext{1}{Astronomy Department, University of Virginia,
P.O. Box 3818, Charlottesville, VA, 22903-0818}

\altaffiltext{2}{Visiting Astronomer, Kitt Peak National Observatory
and Cerro Tololo Inter-American Observatory, National Optical
Astronomy Observatory, which is operated by the Association of
Universities for Research in Astronomy, Inc. (AURA) under cooperative
agreement with the National Science Foundation.}

\altaffiltext{3}{jsilverman@cfa.harvard.edu}
\altaffiltext{4}{pgreen@cfa.harvard.edu}


\begin{abstract}

We present X-ray and optical analysis of 188 AGN identified from 497
hard X-ray (f$_{2.0-8.0 \rm{keV}} > 2.7\times10^{-15}$ erg cm$^{-2}$
s$^{-1}$ ) sources in 20 $Chandra$ fields (1.5 deg$^{2}$) forming part
of the $Chandra$ Multi-wavelength Project.  These medium depth X-ray
observations enable us to detect a representative subset of those
sources responsible for the bulk of the 2--8 keV Cosmic X-ray
Background.  Brighter than our optical spectroscopic limit, we achieve
a reasonable degree of completeness (77\% of X-ray sources with
counter-parts $r^{\prime}<$ 22.5 have been classified): broad emission
line AGN (62\%), narrow emission line galaxies (24\%), absorption line
galaxies (7\%), stars (5\%) or clusters (2\%).  We find that most
X-ray unabsorbed AGN ($N_{\rm{H}}<10^{22}$ cm$^{-2}$) have optical
properties characterized by broad emission lines and blue colors,
similiar to optically-selected quasars from the Sloan Digital Sky
Survey but with a slighly broader color distribution.  However, we
also find a significant population of redder
($g^{\prime}$--$i^{\prime}>1.0$) AGN with broad optical emission
lines.  Most of the X-ray absorbed AGN (10$^{22}<N_{\rm{H}}<10^{24}$
cm$^{-2}$) are associated with narrow emission line galaxies, with red
optical colors characteristically dominated by luminous, early type
galaxy hosts rather than from dust reddening of an AGN.  We also find
a number of atypical AGN; for instance, several luminous AGN show both
strong X-ray absorption ($N_{\rm{H}}>10^{22}$ cm$^{-2}$) and broad
emission lines.  Overall, we find that 81\% of X-ray selected AGN can
be easily interpreted in the context of current AGN unification
models.  Most of the deviations seem to be due to an optical
contribution from the host galaxies of the low luminosity AGN.

\end{abstract}




\section{Introduction}

In the era of $Chandra$ and $XMM-Newton$, X-ray surveys of the
extragalactic universe are for the first time able to probe the
demographics and evolution of the AGN population irrespective of any
moderate obscuration.  Current deep surveys such as the CDF-N
\citep{ba02}, CDF-S \citep{to01} and the Lockman Hole \citep{ma02} are
unveiling both bright quasars and lower luminosity Seyfert galaxies
with significant absorbing gas columns.  This obscuration can be large
enough to effectively hide any optical signature of an active nucleus.
With the unprecedented sensitivity and resolving power of these
current observatories, we are able to probe large volumes to determine
the prevalence of X-ray emitting AGN and their evolution.

The study of AGN enshrouded by dust and gas is not new.  Obscured AGN
(e.g., narrow line radio galaxies, Seyfert 2s, IRAS sources) have been
under investigation for many years, though a complete census of the
population has been out of reach.  The spectrum of the Cosmic X-ray
Background (CXRB) has provided evidence of the preponderance of the
hidden AGN population.  While $ROSAT$ has shown that unabsorbed AGN
dominate the soft (0.1--2 keV) CXRB \citep{ha98}, its high energy
spectrum (2--30 keV) is harder than that of known AGN.  Models based
on the CXRB spectrum and the X-ray luminosity function have predicted
the existence of large numbers of heavily obscured AGN that have been
missed in past surveys \citep{co95,gi01}.

How do these sources fit into the AGN unification scheme (e.g.,
Antonucci 1993; Antonucci \& Miller 1985)?  Many of the absorbed X-ray
sources lack optical AGN signatures (e.g. Barger et al. 2003).  Is
this a result of host dilution \citep{mo02} or some other
geometry/structure that prevents us from viewing the emission line
gas?  While optical extinction and X-ray absorption are statistically
correlated \citep{sm96,tu97}, there are a number of counter examples.
X-ray observations of Seyfert 2 galaxies do not always provide
evidence for a large intrinsic obscuring column (Panessa \& Bassani
2002; Georgantopoulos \& Zezas 2003).  Equally compelling, a number of
X-ray selected type 1 AGN \citep{ma02,ak03} have significant intrinsic
absorption in the X-ray band ($10^{20}<N_{\rm{H}}<10^{23}$ cm$^{-2}$).
\citet{wi02} find that a large fraction of the IR selected AGN found
in the Two-Micron All Sky Survey (2MASS) have broad optical emission
lines and a wide range of X-ray absorption.  Given the complex
environment of some of these AGN, one line of sight might not always
exemplify the overall geometry.

While the $Chandra$ and $XMM-Newton$ deep fields do cover a large volume,
wide field surveys are needed to compile a significant sample of
sources with 2--8 keV flux levels around $10^{-15} - 10^{-14}$ erg
cm$^{-2}$ s$^{-1}$ .  Such sources comprise most of the flux of the
2--8 keV CXRB \citep{mo03,cow02}.  The deep fields provide relatively
few sources at these flux levels to characterize the absorbed AGN
population.  Many ongoing surveys at intermediate flux levels are
currently contributing to our understanding of the X-ray emitting AGN.
For example, the HELLAS2XMM \citep{fi03} and XMM/SSC \citep{bar03}
take advantage of the large field of view and high collecting area of
$XMM-Newton$.  The SEXSI \citep{ha03} survey joins the $Chandra$
Multi-wavelength Project in utilizing $Chandra$'s small PSF and low
background to detect the faint AGN and unambiguously find optical
counter-parts.  With large samples of all AGN types, we can
characterize the dominant population contributing to the CXRB and
determine the relative importance and nature of interesting AGN that
defy a simple unification model.

\section{The $Chandra$ Multi-wavelength Project (ChaMP)}

The ChaMP (Kim et al. 2004a; Kim et al. 2004b; Green et al. 2004) is
providing a medium-depth, wide-area survey of serendipitous X-ray
sources from archival $Chandra$ fields covering $\sim 14$ deg$^2$.  The
broadband sensitivity between 0.3--8.0 keV enables the selection to be
far less affected by absorption than previous optical, UV, or soft
X-ray surveys.  $Chandra$'s small point spread function
($\sim$1$\arcsec$ resolution on-axis) and low background allow sources
to be detected to fainter flux levels, while the $\sim
1^{\prime\prime}$ X-ray astrometry greatly facilitates unambiguous
optical identification of X-ray counter-parts.  The project effectively
bridges the gap between flux limits achieved with the $Chandra$ deep
field observations and those of past $ROSAT$ and $ASCA$ surveys.  A total
of about 8000 serendipitous extragalactic X-ray sources are expected
when the project is complete.  A primary aim of the ChaMP is to
measure the luminosity function of quasars and lower luminosity AGN
out to $z\sim4$ with the inclusion of the obscured population
(J.D. Silverman et al. 2004, in preparation).

We present results from the ChaMP using a subsample
($f_{2.0-8.0\rm{keV}} > 2.7\times10^{-15}$ erg cm$^{-2}$ s$^{-1}$ and
$r^{\prime}<22.5$) of 497 X-ray sources detected in the hard band
(2.5--8.0 keV) in 20 fields.  This work is an extension of the 6
fields analyzed by \citet{gr04}, here limited to the hard X-ray band.
From this subsample, we classify 188 as AGN based on their X-ray
luminosity (L$_{\rm 2-8~kev}>10^{42}$ erg s$^{-1}$).  Our motivation is to
determine the demographics of the hard X-ray emitting AGN, measure the
range of intrinsic obscuration, and determine the extent to which
obscuration of X-rays translates to extinction in the optical.  After
briefly discussing the X-ray and optical data acquisition, reduction
and analysis (\S\ref{hardxray} \& \S\ref{optical}), we describe the
characteristics of the hard X-ray sources (\S\ref{character}) and the
AGN properties (\S\ref{agnsample}) including selection and
completeness .  In section \S\ref{results}, we present the results.
Throughout this paper, we assume H$_{\circ}$=70 km s$^{-1}$
Mpc$^{-1}$, $\Omega_{\Lambda}$=0.7, and $\Omega_{\rm{M}}$=0.3.

\section{X-ray observations}
\label{hardxray}

We have chosen 20 $Chandra$ fields (Table~\ref{xfields}) for which we
have acquired extensive followup optical imaging and spectroscopy.
These fields have been selected from the first two years of $Chandra$
archival data.  Only ACIS observations at high galactic latitude
($|b|>20^{\circ}$) with no special observing modes (e.g., gratings)
are used.  The deepest observations have exposure times that are
sensitive to sources with $f_{2-8\rm{keV}}>2\times10^{-15}$ erg
cm$^{-2}$ s$^{-1}$.  At this flux limit, we resolve $\sim70\%$ of the
2--8 keV CXRB (Moretti et al. 2003; Figure 5).  The target of each
observation has been excluded to avoid any bias towards specific
objects such as AGNs associated with clusters.

A full description of the ChaMP image reduction and analysis pipeline
XPIPE can be found in \citet{ki04a}.  In short, we have an automated
reduction routine that filters out high background intervals, bad
events such as cosmic rays and hot pixels to produce a clean and robust
X-ray source catalog.  Source extraction is performed using a wavelet
detection algorithm (CIAO/{\tt{wavdetect}}; Freeman et al. 2002) in
three energy bands (Broad (B): 0.3--8.0 keV, Soft (S): 0.3--2.5 keV,
Hard (H): 2.5--8.0 keV).

For the following analysis, we require a S/N $>$ 2 in the 2.5--8.0 keV
band to generate a hard X-ray selected sample which minimizes any
inherent bias against the absorption of soft X-rays.  We restrict the
off-axis angle of the detections to less than $12\arcmin$ since the
sensitivity beyond this is significantly reduced.  We do not use chip
S4 (ccd\_id=8) since this CCD is severely affected by a flaw in the
serial readout, causing a significant amount of charge to be randomly
deposited along pixel rows as they are read out.  Each detection has a
unique effective exposure time which includes vignetting.  The
conversion from X-ray count rate to flux units (erg cm$^{-2}$ s$^{-1}$
) is determined from simulated detections on each CCD of a source with
a powerlaw spectrum ($f_{E}\propto E^{-(\Gamma-1)}$;$\Gamma=1.7$\footnote{The ChaMP XPIPE (Kim et
al. 2004a) provides energy conversion factors (ECF) for two models
with $\Gamma=1.7$ and $\Gamma=1.4$.  We chose the former since the
photon index more closely resembles the majority of the hard source
detections.}) and galactic absorption \citep{di90}.  The effect of
varying the photon index ($\Gamma$) from 1.7 to 1.9 results in a
$\sim2\%$ difference in flux for both ACIS-I and ACIS-S.  We calculate
the flux in the conventional 2.0--8.0 keV band for comparison with
other surveys.

With $Chandra$'s broad band sensitivity (0.3--8.0 keV), we are able to
investigate the spectral properties of the sample, though we are
limited by the small number of source counts in most cases (90\% of
sources have $9<{\rm counts}<70$ in the 2.0--8.0 keV band).  The
hardness ratio (HR=H--S/H+S) can be used as a crude assessment of the
spectral characteristics.  Since the response of $Chandra$ varies as a
function of energy and off-axis angle with the additional complication
of mixing frontside and backside illuminated CCDs, we have converted
the raw HR to an effective, on-axis, ACIS-I value by multiplying the
count rate in each band by the ratio of the ECF (off-axis CCD) to the
on-axis ECF which ranges between 0.6--2.0.

\subsection{X-ray spectral fits}

\label{xfit}

X-ray spectral modeling provides a robust way of characterizing the
spectral properties of our sample, independent of observation and
instrument details.  With a measured redshift, we can more accurately
determine the intrinsic absorbing column than that based solely on
hardness ratios.  Some objects that look soft in HR may have
significant absorption especially at higher redshifts.  For each X-ray
source in our hard-selected sample, we use an automated procedure to
extract the spectrum and fit a model to the data.  Due to a lack of
counts, we cannot usefully fit a spectral powerlaw model, leaving both
spectral index and intrinsic absorbing column free for all objects.

All processing is done using CIAO
3.0.2\footnote{http://cxc.harvard.edu/ciao} and CALDB
2.26\footnote{http://cxc.harvard.edu/caldb}.  The detailed steps to
prepare the PHA (Pulse-Height Analysis) spectrum follow.  First, we
define a circular region centered on the X-ray source sized to contain
95\% of 1.5\,keV photons at the given off-axis angle.  The background
region is annular with a width of $20\arcsec$ centered on the
source. We exclude any nearby sources from both the source and
background regions.  We then use CIAO tool {\em psextract} to create a
PHA spectrum covering the energy range 0.4--8\,keV.  We generate both
an ungrouped spectrum and one which is grouped to a minimum of 10
counts per channel.  The time-dependent quantum efficiency degradation
of ACIS is accounted for when the ARF is generated by the {\em mkarf}
tool.

Spectral fitting is done using the CIAO {\em
Sherpa}\footnote{http://cxc.harvard.edu/sherpa} tool.  For all
sources, we fit an absorbed powerlaw containing an intrinsic absorber
with neutral column $N_{\rm{H}}$ at the source redshift.  Our choice
of photon index (Frozen at $\Gamma=1.9$) is based on previous studies
of unabsorbed AGN.  \citet{re00} have measured the spectral index for
radio quiet AGN using $ASCA$ observations to be $\Gamma\sim1.9$.
\citet{pi03} have measured a mean photon index 1.8--1.9 with
$XMM-Newton$ which shows no variation over the redshift range $0<z<2$.
This $N_{\rm H}$ fit provides a robust one-parameter characterization
of the intrinsic spectral shape for as few as 10 counts.  We verified
by an extensive Monte-Carlo simulation that the parameter
uncertainties calculated with {\em projection} of confidence contours
in {\em Sherpa} are reliable. Note that the spectral model contains a
fixed Galactic neutral absorber appropriate for each object.  Spectra
with at least 60 counts are fit using the grouped spectrum with the
hybrid Monte-Carlo Levenberg-Marquardt minimization method, while the
low count spectra are fit using the ungrouped data with Cash
statistics and the Powell method.  Spectra with over 200 counts are
also fit with a two-parameter absorbed powerlaw leaving both $\Gamma$
and the intrinsic $N_{\rm{H}}$ at the source redshift free to vary.
The results from the two-parameter fitting are included in this paper
for a couple of sources discussed in Section~\ref{unabsorbed_nolines}.
The full analysis will be presented in an upcoming ChaMP paper
(T.L. Aldcroft et al. 2004, in preparation).

\section{Optical followup}

\label{optical}

\subsection{Imaging}

We have acquired optical imaging for each $Chandra$ field to identify
counter-parts to X-ray sources.  We use the NOAO Blanco and Mayall 4m
telescopes and their MOSAIC cameras to image the full $Chandra$
field-of-view of our survey fields.  The exposure times are scaled
from the minimum X-ray flux for a detection per $Chandra$ field to
identify $>90\%$ of $ROSAT$ AGN \citep{yu98}.  Three filters ($g^{\prime}$,
$r^{\prime}$, $i^{\prime}$) using the Sloan Digital Sky Survey (SDSS)
photometric system \citep{fu96} are implemented to measure broad band
colors for preliminary source classification.  Table 2 provides some
details of the imaging for the fourteen fields not included in Green
et al. (2004).

A full description of the optical followup program including strategy,
image reduction, source detection and photometric calibration can be
found in \citep{gr04}.  Briefly, optical image reduction on the MOSAIC
data is performed with the {\tt mscred} \citep{valdes02} package
within the IRAF\footnote{IRAF is distributed by the National Optical
Astronomy Observatory, which is operated by the Association of
Universities for Research in Astronomy, Inc., under cooperative
agreement with the National Science Foundation.} environment.  We use
the SExtractor \citep{be96} algorithm to detect sources, and measure
their positions and brightness.  Since the X-ray source positions are
only accurate to within $\sim1\arcsec$, the optical astrometric
solution is required to achieve the rms $<0.3\arcsec$ accuracy
necessary for spectroscopic followup.  We require an accuracy of the
photometric solution to less than a tenth of a magnitude.  The
majority of the AGN sample (88\%) presented in this paper has
magnitude errors less than 0.05 as a result of the bright optical
magnitude selection.  In addition, we find a mean color offset of 0.05
magnitudes to the red between our $g^{\prime}-i^{\prime}$ color and
the SDSS using sources detected in both surveys.

\subsection{X-ray to optical source matching}

\begin{figure}
\epsscale{1.0}
\plotone{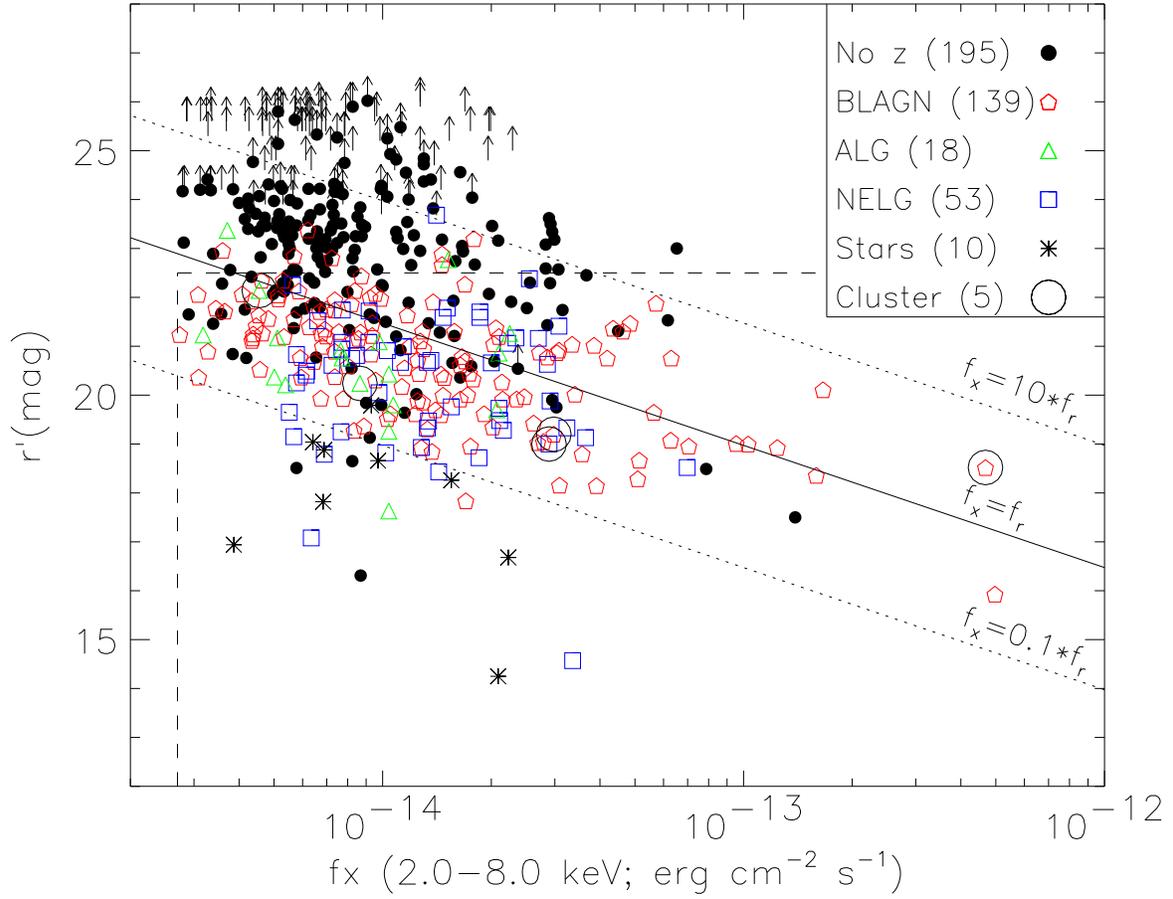}

\caption{X-ray flux (2--8 keV) vs. optical magnitude ($r^{\prime}$).
Optical spectroscopic classifications are indicated (top right box)
with the sample size in parenthesis.  X-ray sources with no optical
counter-parts are shown by an arrow placed at the hypothetical
magnitude for a 5$\sigma$ detection from our optical imaging (Table
2).  The dashed vertical and horizontal lines mark the X-ray flux
limit and optical magnitude limit for the subsequent analysis.  The
slanted lines mark the f$_{\rm{X}}$/f$_{\rm{r}}$ ratios of 0.1, 1, 10.}
\label{fxopt}
\end{figure}

We implement an automated routine to match each X-ray source with
potential optical counter-parts (see Green et al. 2004).  The search
radius is increased for X-ray detections at large off-axis angles.
Each match is visually inspected and a confidence level is determined.
For X-ray sources with multiple optical counter-parts, the optical
source closest to the X-ray centroid is usually given a higher
confidence.  We have found 415 optical counter-parts to 497 X-ray
sources (84\%; Figure~\ref{fxopt}).  We have not included 36 sources
because the X-ray detection fell on a chip edge or there were multiple
optical counter-parts for which no single optical source could
confidently be assigned.  We note that 81\% of the matches have an
X-ray to optical offset $<2\arcsec$.

\subsection{Spectroscopy}

Optical spectroscopy is crucial for determining the source type and
redshift.  We acquired the majority of our optical spectra with the
WIYN/3.5m and CTIO/4m with the HYDRA multi-fiber spectrographs, which
have a field of view ($>40\arcmin$) that fully covers the $Chandra$
field.  To extend spectroscopic classifications beyond
$r^{\prime}\sim21$, the limit of the 4m class telescopes with HYDRA,
we have obtained spectra from Magellan and the MMT.  The field of view
of Magellan with LDSS-2, a multi-slit spectrograph, is 5\arcmin, so it
takes 5--6 pointings to cover the full $Chandra$ field.  We have been
using the FLWO 1.5m to acquire spectra of the optically bright
($r^{\prime}<17$) counter-parts.  In addition, a number of people,
mentioned in the acknowledgements, have graciously acquired long slit
spectra of a few ChaMP sources during their own observing time.  All
redshifts have an accuracy of $\Delta z<0.001$.  Table~\ref{ospec}
gives a summary of the spectroscopic facilities used by the ChaMP
project.

We implement a classification scheme of optical spectra similar to the
{\em Einstein Observatory} Extended Medium-Sensitivity Survey (Stocke
et al. 1991).  Objects with strong emission lines (W$_{\lambda}>5$
\rm{${\rm \AA}$}) are classified as either Broad Line AGN (BLAGN; FWHM
$>$ 1000 km s$^{-1}$) or Narrow Emission Line Galaxy (NELG; FWHM $<$
1000 km s$^{-1}$).  Counter-parts with weak emission line or pure
absorption line spectra are classified as Absorption Line Galaxy
(ALG).  We note that there is a combination of redshift range and
spectral bandpass for which we may lose important AGN optical
diagnostic features.  In some cases, the host galaxy contribution can
prevent H$\beta$ from being a useful AGN indicator.  Given the low
signal-to-noise of our spectra, some NELG may have weak, broad
emission lines.  \citet{ho97} found that broad H$\alpha$ can often be
found in low luminosity ``dwarf'' Seyferts with high S/N spectra and
proper subtraction of the stellar continuum.  This type of analysis is
not possible given the quality of our spectra.  Any stellar source is
labelled as a STAR.  For the ALG, we measure the Ca II break
``CONTRAST" (Stocke et al. 1991) to look for a powerlaw AGN component
to the continuum to note potential BL LAC candidates.  If the
associated X-ray emission is extended the object is further labelled
as a possible cluster member.

As shown in Figure~\ref{fxopt}, we have classified 44\% (220) of all
the hard X-ray sources through our spectroscopic campaign.  The
sources without redshifts are primarily at faint optical magnitudes
($r^{\prime}>22$).  In Table~\ref{numbers}, we list the numbers of
each type for various limits imposed on the sample.

\section{Characteristics of the hard X-ray sample}
\label{character}

\subsection{X-ray and optical flux}

We show the optical magnitude ($r^{\prime}$) as a function of X-ray
flux (2.0--8.0 keV) for the 497 sources detected in 20 fields
(Figure~\ref{fxopt}).  Lines of constant
$f_{\rm{X}}$/f$_{r^{\prime}}$ are determined as follows.
\begin{equation}
log(f_{\rm{X}}/f_{r^{\prime}})=log(f_{\rm X})+0.4r^{\prime}+5.41
\end{equation}
This relation has been derived using an assumed powerlaw ($f_{E}\propto
E^{-\alpha}$) with spectral index $\alpha_{O}=0.5$ and
$\alpha_{X}=0.7$. The characteristics of the $r^{\prime}$ filter are
taken from \citet{fu96}.  Most objects have $0.1<$
$f_{\rm{X}}/f_{r^{\prime}}<10$.  There exists a significant
number of X-ray bright, optically faint sources
(f$_{\rm{X}}/f_{r^{\prime}}>10$), many of which are not detected
in our optical imaging.  Due to their optical faintness, we have only
identified one such source, an NELG.

Based on the 210 spectroscopically identified objects with
$r^{\prime}<$ 22.5, we find that 62\% of the hard X-ray sources are
classified as BLAGN.  As shown in \citet{gr04}, these AGN tend to
follow the relation $f_{\rm{X}}=f_{r^{\prime}}$ over a wide range
of optical and X-ray flux.  The NELG, which comprise 24\% of the
identifications, have flux ratios similar to the BLAGN.  We find a
number of counter-parts (7\% ALG) that have no evidence for an
emitting line region.  These galaxies are primarily identified at
bright optical magnitudes ($r^{\prime}<21$) due to the difficulty of
classifying sources without strong emission lines at high redshift
($z>0.5$).  In addition, a few hard X-ray detected sources are
associated with optically bright stars (5\%) and galaxy clusters (2\%)
with extended X-ray emission.

\subsection{X-ray spectral properties}

\label{text:hardhr}

\begin{figure}
\epsscale{1.0}
\plotone{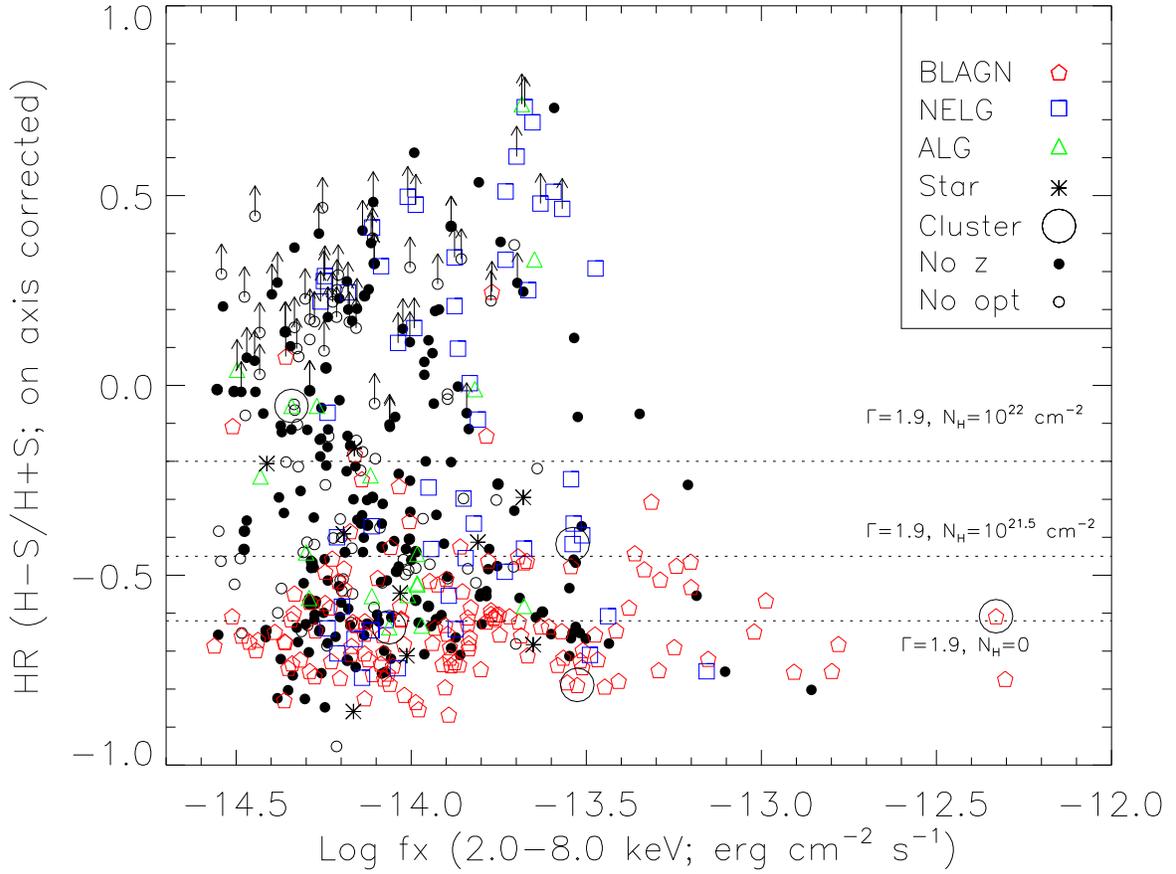}

\caption{X-ray hardness ratio (HR) as a function of 2.0--8.0 keV flux.  The
hardness ratio has been recalculated to the equivalent on-axis value
for all sources.  Error bars have been omitted for clarity and lower
limits are marked by an arrow. Symbol type indicates the spectroscopic
classification.  Objects with no spectrum are marked as filled
circles, while objects with no optical counter-part are marked by a
small open circle.}
\label{hr}
\end{figure}

The X-ray hardness ratio provides a crude measure of the spectral
properties and classification of the hard X-ray sources.  In
Figure~\ref{hr}, we plot the corrected hardness ratio
(Section~\ref{hardxray}) as a function of X-ray flux.  The flux range
shown includes all sources with the exception of the extremely bright
cataclysmic variable TX Col \citep{sc04}.  The horizontal lines mark
the hardness ratio which corresponds to an X-ray source with a
powerlaw continuum (photon index $\Gamma=1.9$) absorbed by a column of
gas at $z=0$.  Some of the derived $N_{\rm H}$ detections may be
higher (see Section~\ref{xfit}) using an absorber intrinsic to the
source.  According to Moretti et al. (2003), we are resolving
$\sim70\%$ of the full 2--8 keV CXRB at our chosen flux limit.  With a
flux weighted mean HR for the ChaMP sources of --0.39, the X-ray
spectrum of the ensemble is similar to the spectral characteristic of
the integrated CXRB ($\Gamma$=1.4) which corresponds to a hardness
ratio of --0.42.

As described in many studies with $Chandra$ (e.g., Mushotzky et
al. 2000) and $XMM-Newton$ (e.g., Hasinger et al. 2001), the X-ray
source population becomes relatively harder at fainter flux levels.
As evident in Figure~\ref{hr}, the hardest (HR$>$0) X-ray sources in
these ChaMP fields are predominately at log$_{10}$($f_{\rm X}$)$<$
--13.6.  At fainter flux levels, the sources have a more even
distribution over all hardness ratios.  This spectral variation has
been attributed to intrinsic absorption rather than changes in the
intrinsic spectral energy distribution (Mainieri et al. 2003; Kim et
al. 2004b).  This will be further investigated in an upcoming ChaMP
paper on the X-ray spectral properties of the AGN sample.

\section{Hard AGN sample}

\label{agnsample}

\subsection{Selection and completeness}

\label{selection}

In these medium depth $Chandra$ fields, we are sensitive to X-ray
sources with a 2.0--8.0 keV flux greater than $2.7\times10^{-15}$ erg
cm$^{-2}$ s$^{-1}$.  The addition of a spectroscopic limit of
$r^{\prime}<22.5$, yields a sample of X-ray sources which is 77\%
identified.  This optical magnitude limit does bias the sample against
optically faint counter-parts at the lower X-ray flux levels.

To construct a pure AGN sample, we require the rest frame 2.0--8.0 keV
luminosity (uncorrected for intrinsic absorption) to exceed 10$^{42}$
erg s$^{-1}$ thereby excluding any sources that may contain a
significant stellar or hot ISM component.  The most luminous known
star-forming \citep{ze03,li02} or elliptical \citep{os01} galaxies
attain at most $L_{\rm X}=10^{42}$ erg s$^{-1}$.  Since many of the
traditional optical AGN signatures are not present in obscured
sources, high X-ray luminosity becomes our single discriminant for
supermassive black hole accretion.  We believe that almost all of the
NELG and ALG harbor accreting SMBHs based on their X-ray luminosity.
We find that 90\% of the identified ChaMP sources have luminosities
above this threshold.  These selection criteria yield a sample of 188
AGN from 20 $Chandra$ fields with $f_{\rm 2-8 keV}>2.7\times10^{-15}$
erg cm$^{-2}$ s$^{-1}$, $r^{\prime}<22.5$ and $L_{\rm X}>10^{42}$ erg
s$^{-1}$.  We have removed 5 objects identified as clusters based on
their extended X-ray emission.  A truncated version of the AGN catalog
is presented in Table~\ref{catalog} and a full version is available in
electronic form.  This sample is composed of 69\% BLAGN, 24\% NELG and
7\% ALG (Table~\ref{numbers}).

\subsection{Redshift and luminosity distribution}

\begin{figure}
\epsscale{1.1}
\plottwo{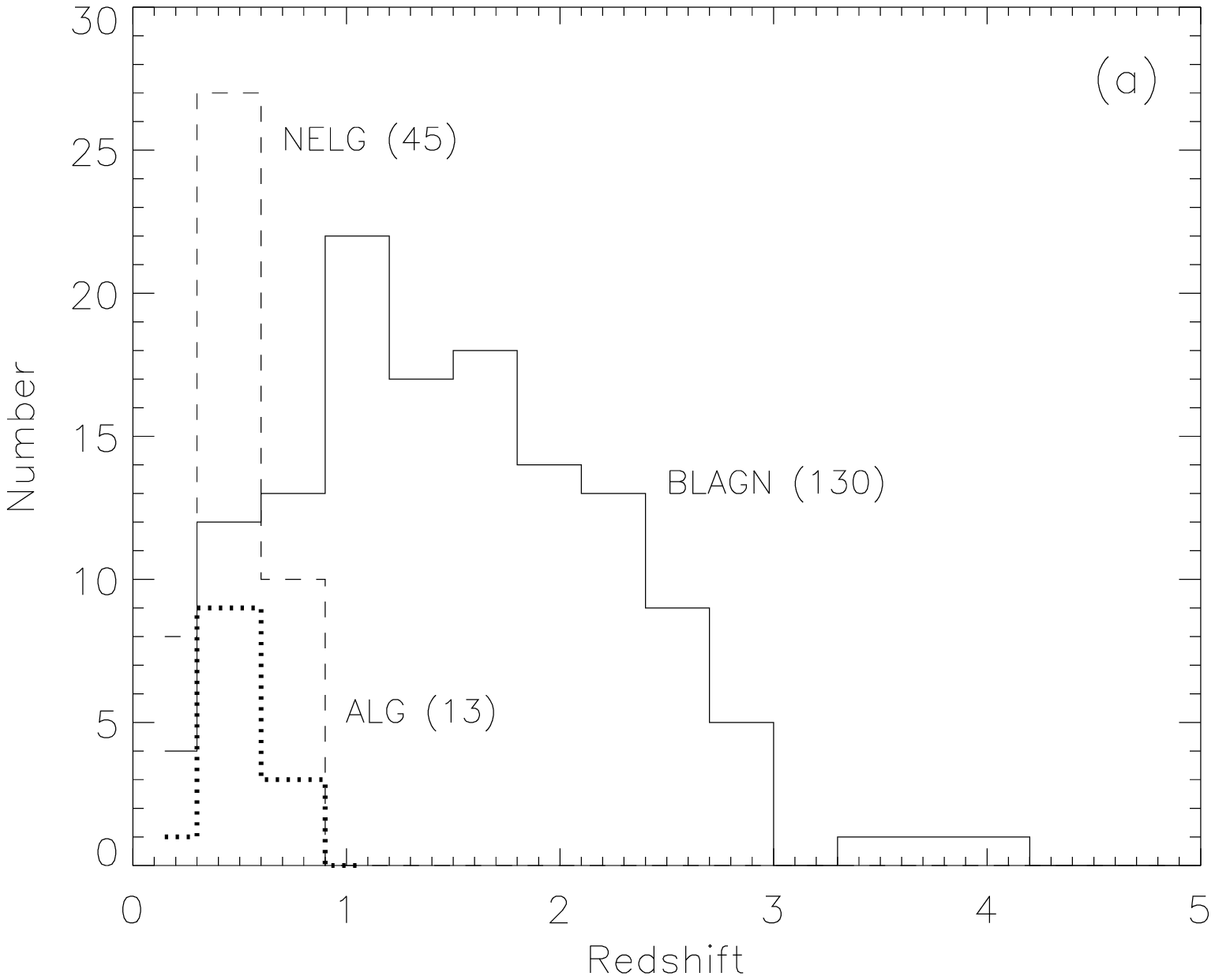}{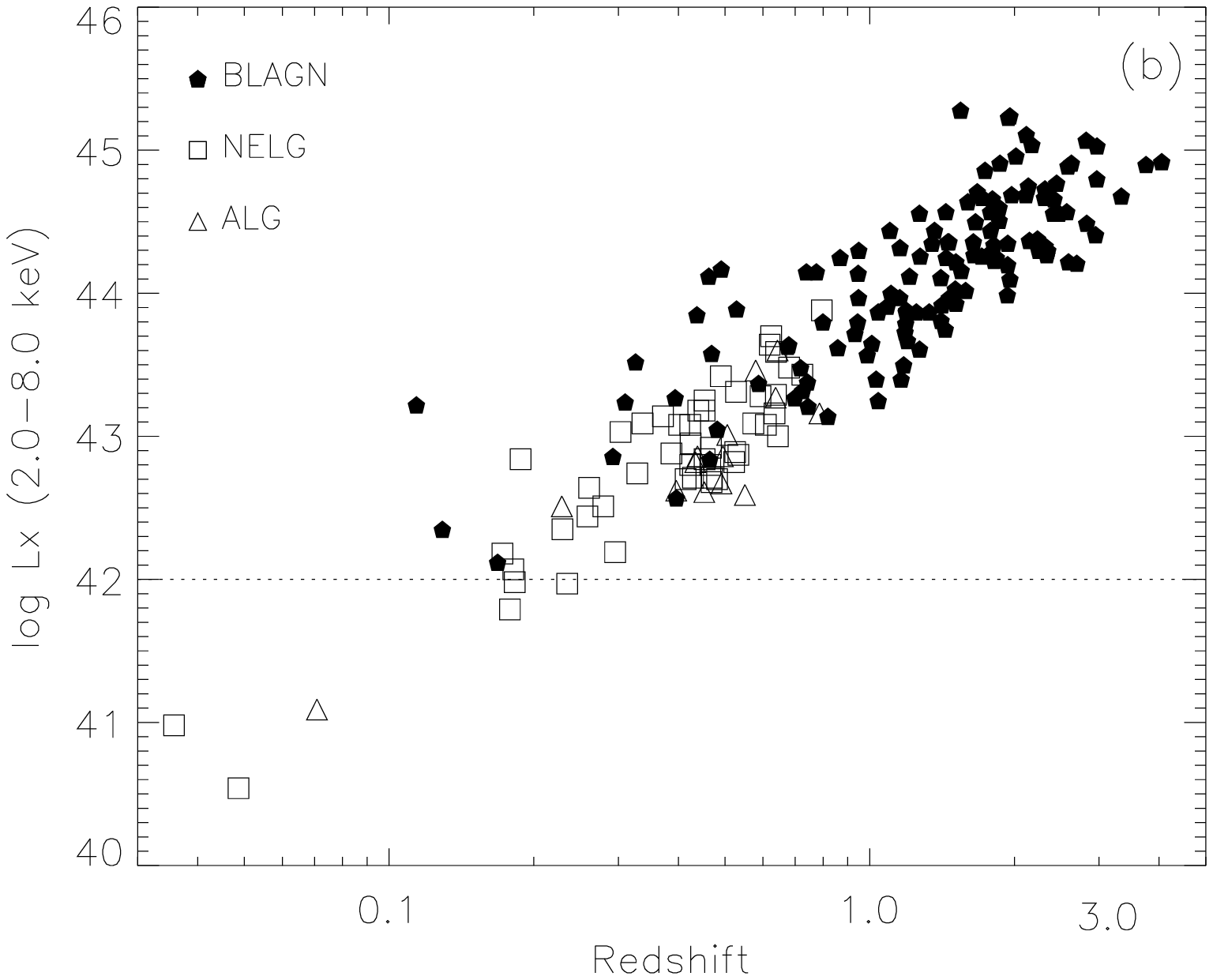}

\caption{(a) Redshift distribution of hard band detected AGN with
L$_{2.0-8.0 \rm{keV}} > 10^{42}$ erg s$^{-1}$.  (b) X-ray luminosity,
redshift distribution.  The horizontal, dashed line marks our chosen
minimum luminosity required for AGN selection.  All but one NELG
($z=0.014$; log (L$_{2.0-8.0 \rm{keV}}$)= 39.5) are shown.}
\label{lzdistr}
\end{figure}

As shown in Figure~\ref{lzdistr}, $Chandra$ detects hard BLAGN out to
$z\sim$ 4 in these medium depth observations.  In contrast, we have
only identified non-BLAGN with $z<0.8$.  The steep drop in the number
of NELG and ALG above this redshift is primarily due to an optical
selection bias.  The NELG and ALG lack prominent broad emission lines
and may be resolved, having a magnitude fainter than our
$r^{\prime}$=22.5 limit within our spectroscopic aperture.  Even a
luminous host galaxy (10L$_{\star}$) at $z\sim0.8$ is fainter than
this limit.  However, the $Chandra$ Deep field surveys are finding a
peak in the number distribution at $z\sim0.8$ with a sharp dropoff
beyond $z\sim1$ \citep{cow03}.  Therefore, the peak in our
distribution for NELG and ALG is unlikely to change drastically with
the identification of the optically faint sources.

\subsection{Low luminosity AGN and composite systems}
\label{dilution}

Even though we have mitigated bias against obscured AGN by carrying
out a hard X-ray survey, we still need to incorporate the optical
properties of these AGN to determine their nature and how they fit
into the unification scheme.  Many of the newly detected $Chandra$ and
$XMM-Newton$ sources in the deep fields show no evidence for an AGN in
the optical.

\begin{figure}
\epsscale{1.0}
\plotone{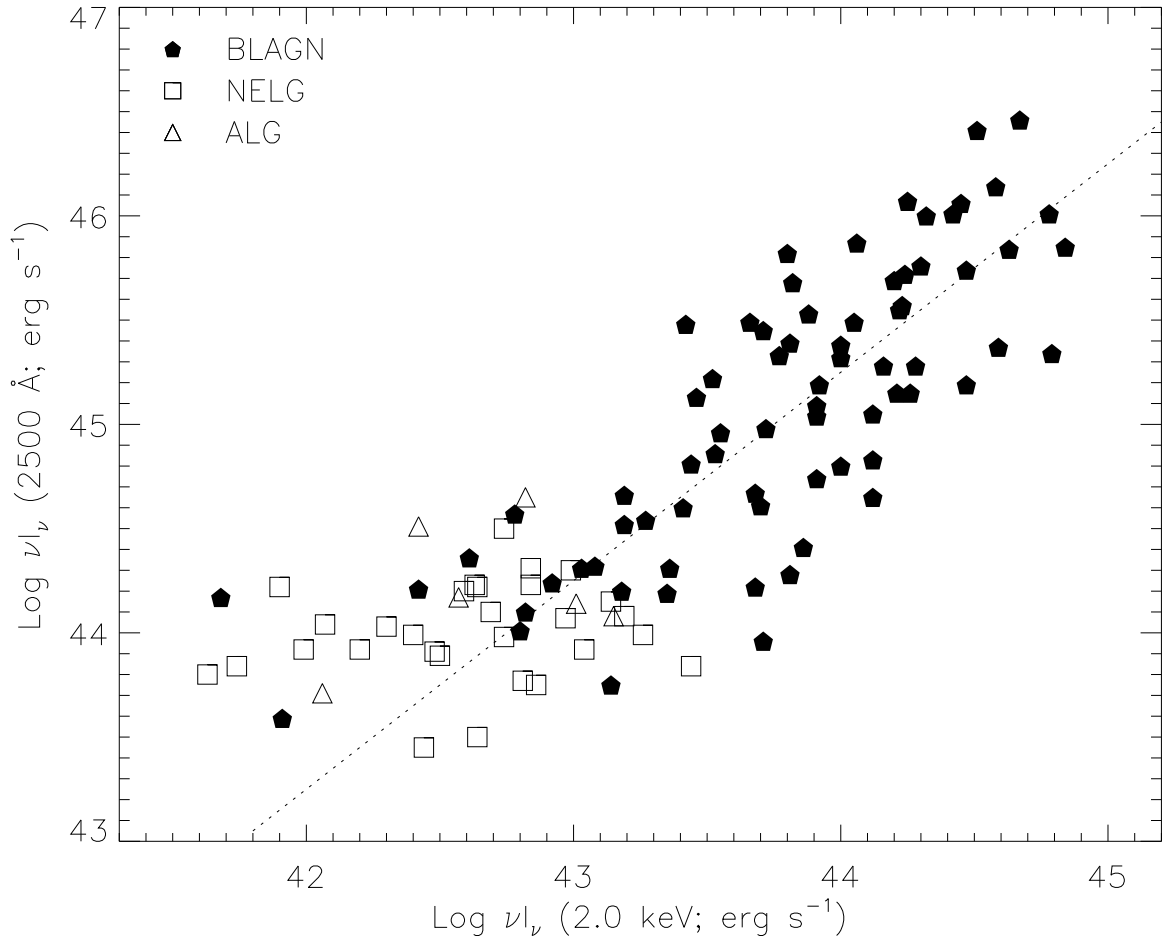}

\caption{X-ray versus optical luminosity.  The dashed line is the mean
X-ray to optical flux ratio ($<\alpha_{OX}>=1.48$) of the BLAGN with
log L$_{2.0-8.0 \rm{keV}} > 43.5$ (units of erg s$^{-1}$).
Propogating typical errors in a Monte-Carlo simulation for each object
yields error bars of similar size to the points shown here.}
\label{lxlo}
\end{figure}

Optical emission from these low luminosity AGN can include a
significant contribution from the host galaxy (Fiore et al. 2003;
Green et al. 2004).  This tends to be the case for AGN with
$L_{X}<10^{43}$ erg s$^{-1}$.  We plot in Figure~\ref{lxlo} the
monochromatic, rest frame optical (2500 ${\rm \AA}$) versus X-ray (2
keV) luminosity density for our hard AGN sample with $f_{2-8~{\rm
keV}}>1\times10^{-14}$ erg cm$^{-2}$ s$^{-1}$ and $r^{\prime}<22.5$.
We measure the optical luminosity density at 2500 ${\rm \AA}$ from the
$r^{\prime}$ magnitude as in Fukugita et al. (1996), assuming a
powerlaw spectrum with $\alpha_{O}=0.5$.  The monochromatic X-ray
luminosity is derived from the hard band flux using a powerlaw
spectrum with $\alpha_{X}=1.0$.  The additional X-ray flux limit is
imposed to include BLAGN across the full range of $\alpha_{OX}$
($\sim1-2$).  If we included BLAGN with fainter X-ray fluxes, we
would be biasing the sample too strongly towards optically bright
objects based on our optical magnitude limit.  The BLAGN above log
$\nu l_{\nu}(\rm 2.0~keV)> 43.5$ share a similar X-ray to optical
ratio ($<\alpha_{OX}>=1.48\pm0.03$) while the lower luminosity AGN
depart from this relation.

\begin{figure}
\hspace{-1cm}
\epsscale{1.2}
\plottwo{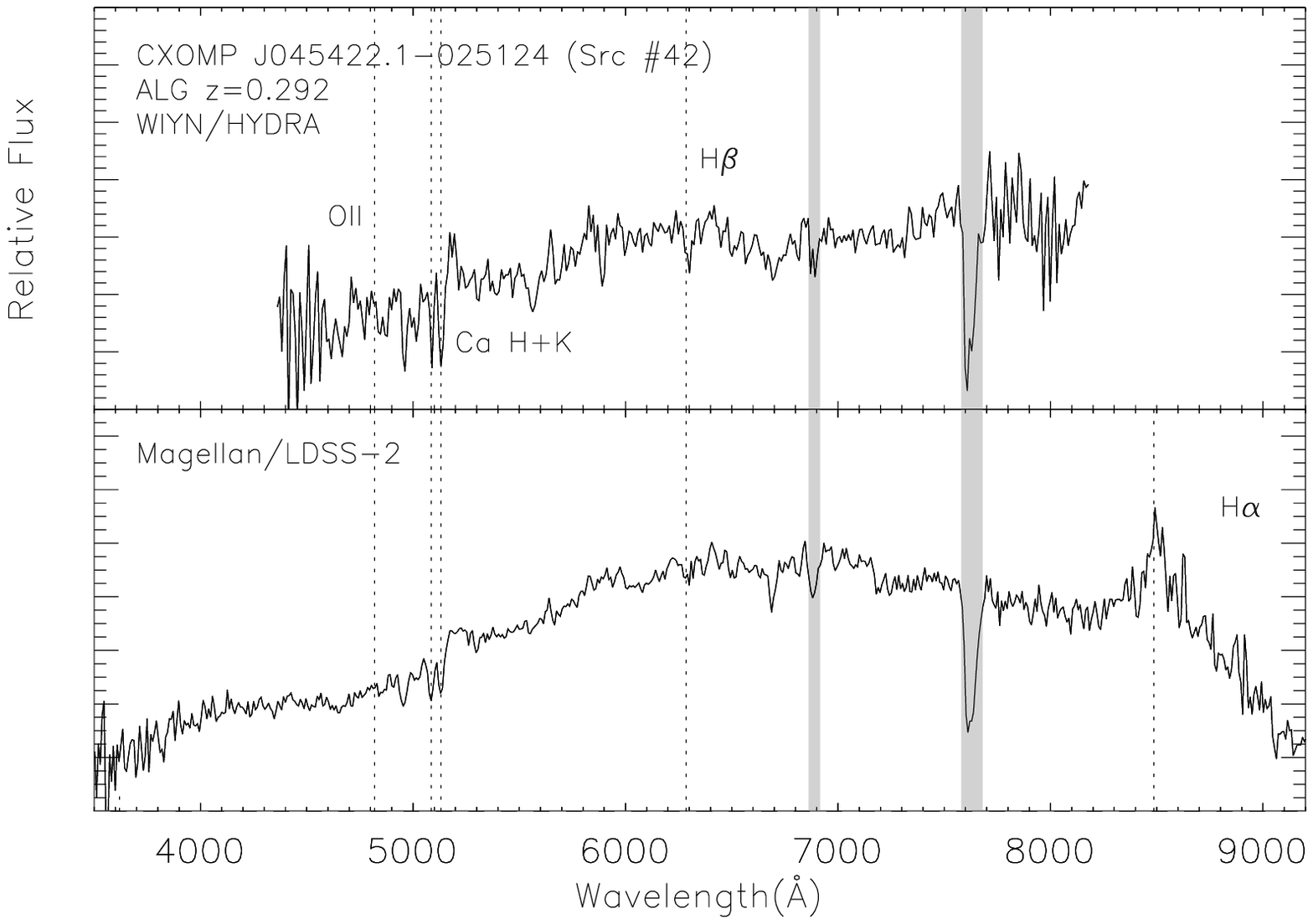}{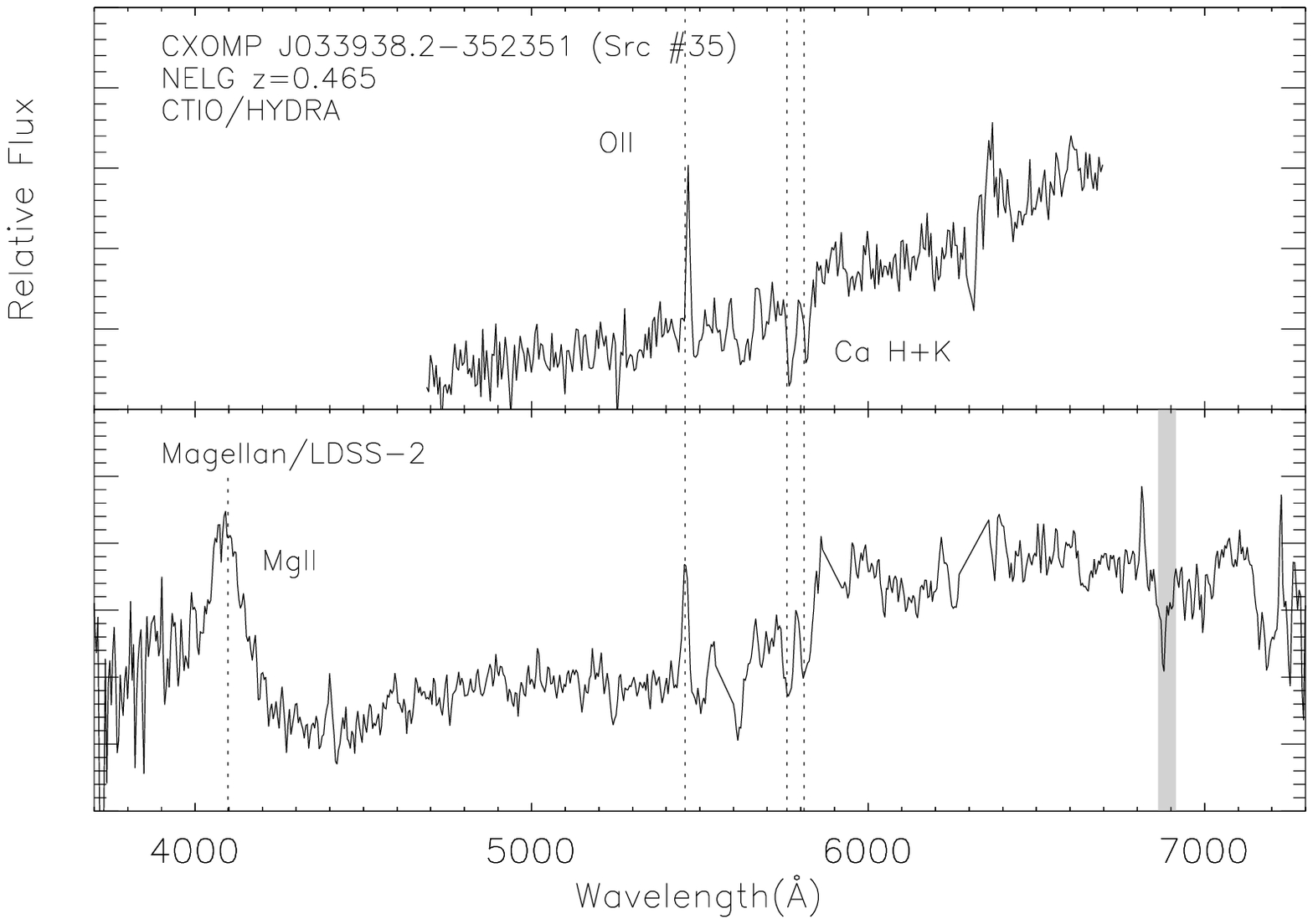}

\caption{Complexity of object classification.  Two optical spectra of
each source were acquired with slightly different wavelength coverage.
The spectra taken with Magellan clearly reveal a broad emission line
in each case.  Shaded regions mark the uncorrected telluric O$_{2}$
absorption features.  Dashed lines mark the expected observed
wavelengths of emission or absorption features at the source
redshift.}
\label{misclass}
\end{figure}

Most studies of X-ray selected AGN rely on optical spectroscopy for
further classification.  This can be misleading since it is difficult
to isolate the nuclear region at these distances (Moran, Filippenko \&
Chornock 2002) with a $\sim1\arcsec$ aperture.  While most of the
optical counter-parts of ChaMP sources have emission lines similar to
quasars and lower luminosity Seyfert galaxies, limited wavelength
coverage can cause some confusion with source classification as
described by \citet{pa03}.  For example, we display in
Figure~\ref{misclass} two sources each with optical spectra spanning
different wavelength ranges.  Depending on the observed wavelength
coverage, the classification can change drastically with the detection
of broad emission lines either at the blue end of the spectrum (Src \#
35; Mg II) or the red end (Src \#42; H$\alpha$).  However, the
inclusion of these AGN in our sample highlight the power of X-ray
selection to reveal supermassive black hole accretion even under a
veil of obscuration or host contamination.

\subsection{Comparison with the optically selected samples (SDSS)}

\begin{figure}

\hspace{-1cm}
\epsscale{0.6}
\plotone{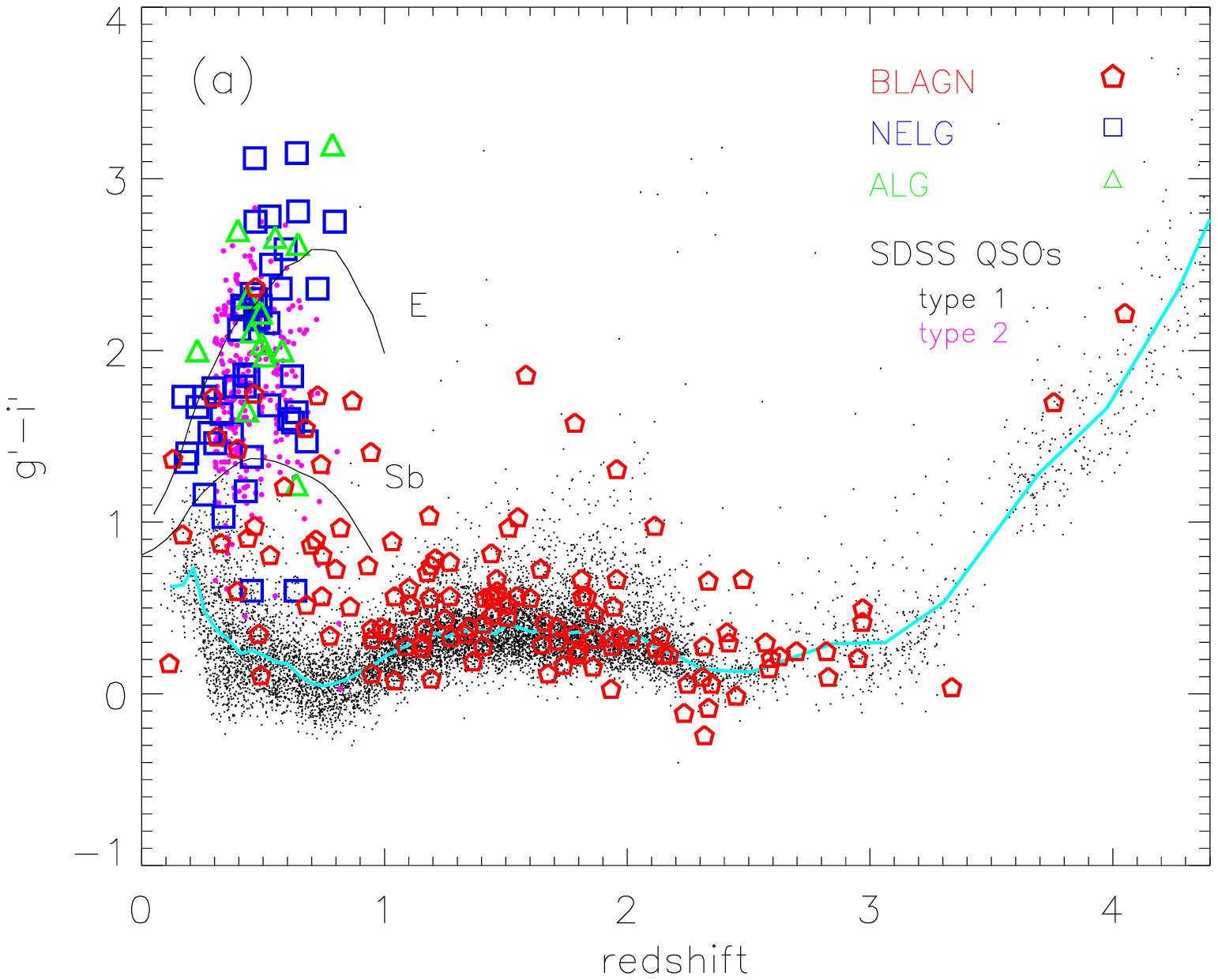}
\plotone{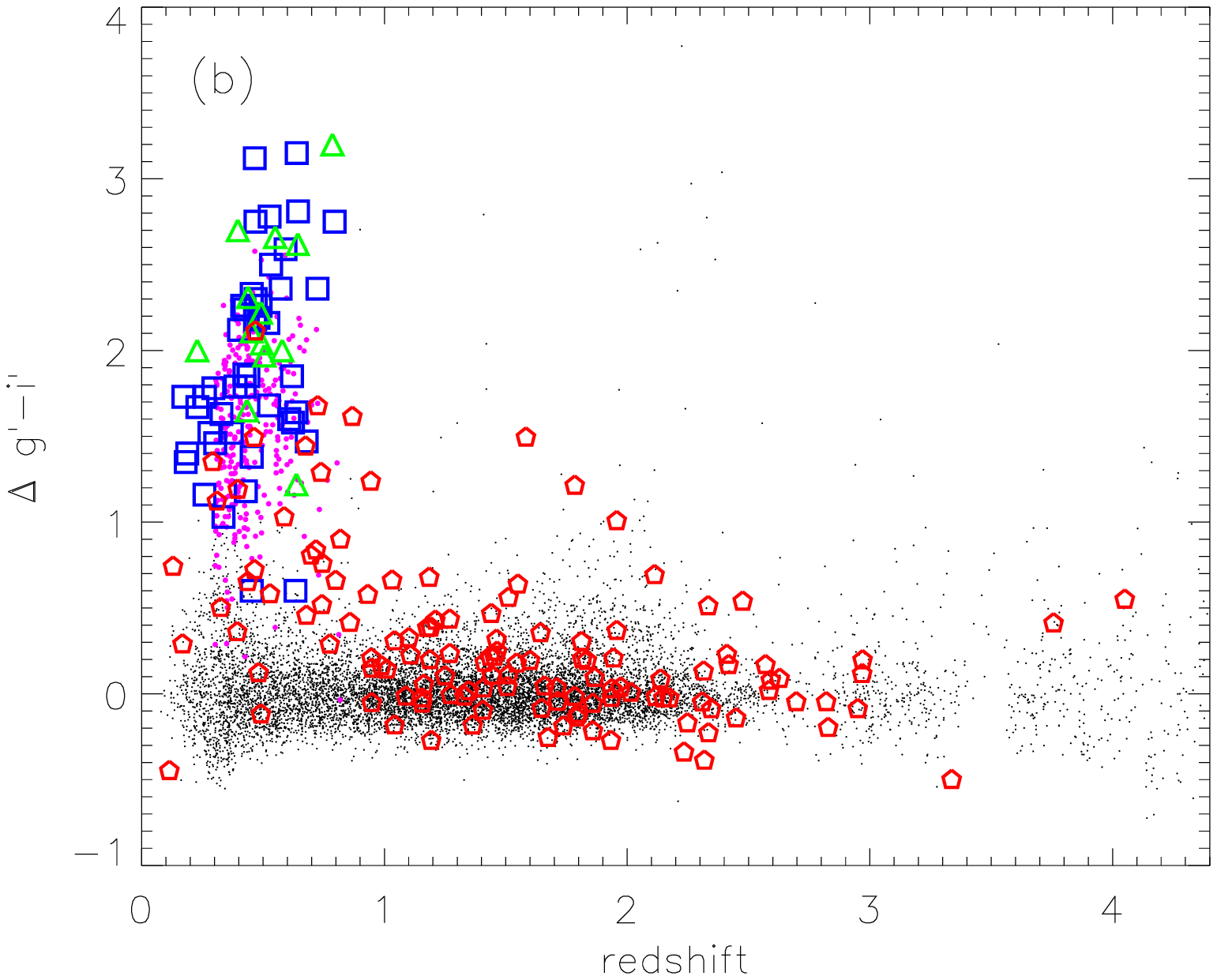}
\caption{(a) Optical color of ChaMP hard AGN compared with type 1
(Schneider et al. 2003) and type 2 (Zakamska et al. 2003) SDSS
quasars.  The median type 1 SDSS quasar color in redshift bins of 0.05
($z<2.2$), 0.2 ($2.2<z<2.6$) and 0.3 ($z>2.6$) is shown by the cyan
curve.  (b) Color offsets from the median type 1 quasar color from the
SDSS.}
\label{sdss}
\end{figure}

With optical imaging in three SDSS bands ($g^{\prime}$, $r^{\prime}$,
$i^{\prime}$), the SDSS becomes the easiest and largest sample of AGN
to which we can compare our ChaMP AGN.  The second edition of the SDSS
quasar catalog \citep{sc03} contains 16,713 objects with $M_{i^\prime}<$--22
out to $z=5.41$ from the first public data release.  We have selected
10,736 of these which were targeted as quasars solely based on their
optical colors (BEST spectroscopic flag=1 for either low-z and high-z
quasars).  In Figure~\ref{sdss}a, we have plotted the optical color
($g^{\prime}$--$i^{\prime}$) of the ChaMP AGN and the quasars from the
SDSS as a function of redshift.  Following \citet{ri01}, we have
measured the color offset (Figure~\ref{sdss}b) from the median color
of type 1 quasars from the SDSS.      

\begin{figure}

\epsscale{0.9}
\plotone{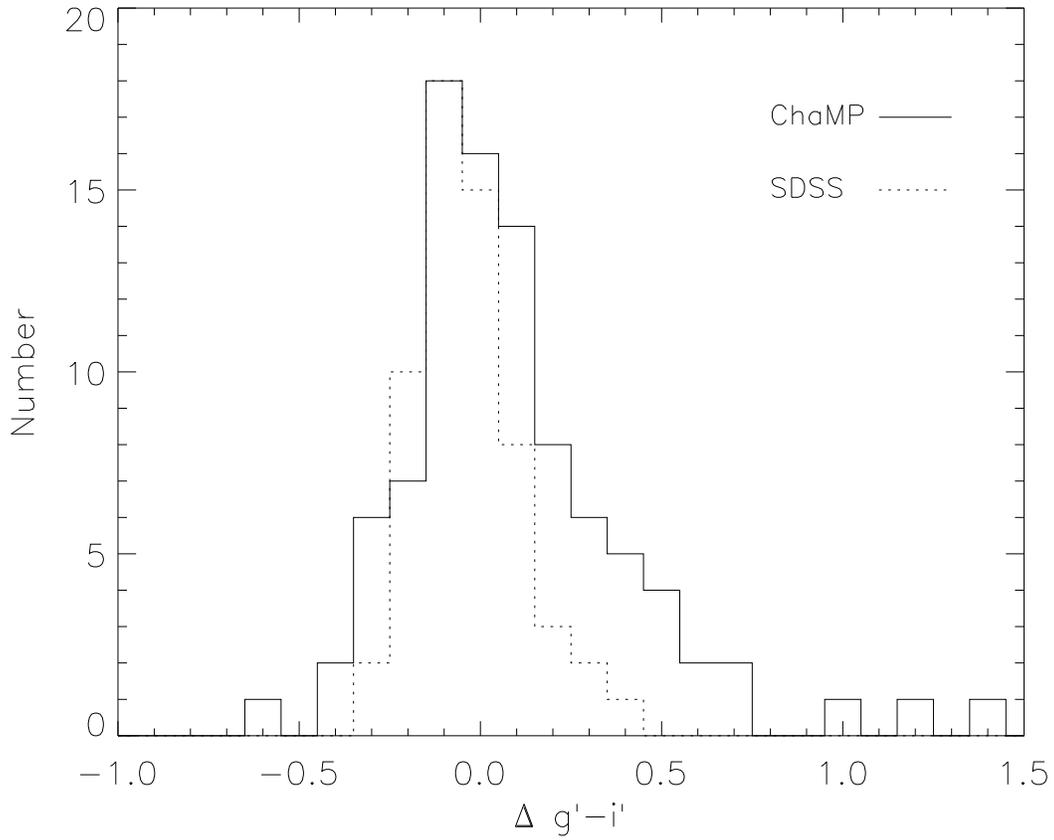}

\caption{Color difference of ChaMP BLAGN and SDSS quasars with the
median SDSS quasar color.  A minimum redshift ($z>1$) has been chosen
to omit low luminosity AGN with a significant host component.  The
SDSS distribution has been normalized to match the peak of the ChaMP
AGN.}
\label{color_comp}
\end{figure}

The ChaMP BLAGN follow the color locus of SDSS quasars
(Figure~\ref{sdss}) for $z>1$.  The ChaMP BLAGN are slightly redder
with a color excess of 0.1 magnitudes compared to the median SDSS
quasar colors (Figure~\ref{color_comp}).  Given the photometric
accuracy (0.1 mag) of ChaMP, we cannot with strong confidence report a
difference of the mean between the two samples, though we probably
achieve a higher precision with our sample of 95 ChaMP BLAGN.  We do
find a 0.03\% likelihood that the color distribution of the ChaMP
BLAGN ($z>1$; $M_{i^{\prime}}<$ --22) can be drawn from the SDSS quasar
population using a Kolmogorov-Smirnov test (Press et al. 1992).  In
Figure~\ref{color_comp}, we see that the ChaMP BLAGN have a wider
color distribution ($\sigma_{\Delta g^{\prime}-i^{\prime}}$=0.41) than
the SDSS quasars ($\sigma_{\Delta g^{\prime}-i^{\prime}}$=0.26).
Besides the width of the distribution, we can compare the symmetry
about the mean by measuring the skewness of the distribution, defined
in Equation~\ref{eq:skewness}.

\begin{equation}
skewness=\frac{\sum_{i=1}^{N}(y_{i}-\overline{y})^{3}}{(N-1) \sigma^3}
\label{eq:skewness}
\end{equation}  

\begin{figure}

\hspace{-1cm}
\epsscale{1.1}
\plottwo{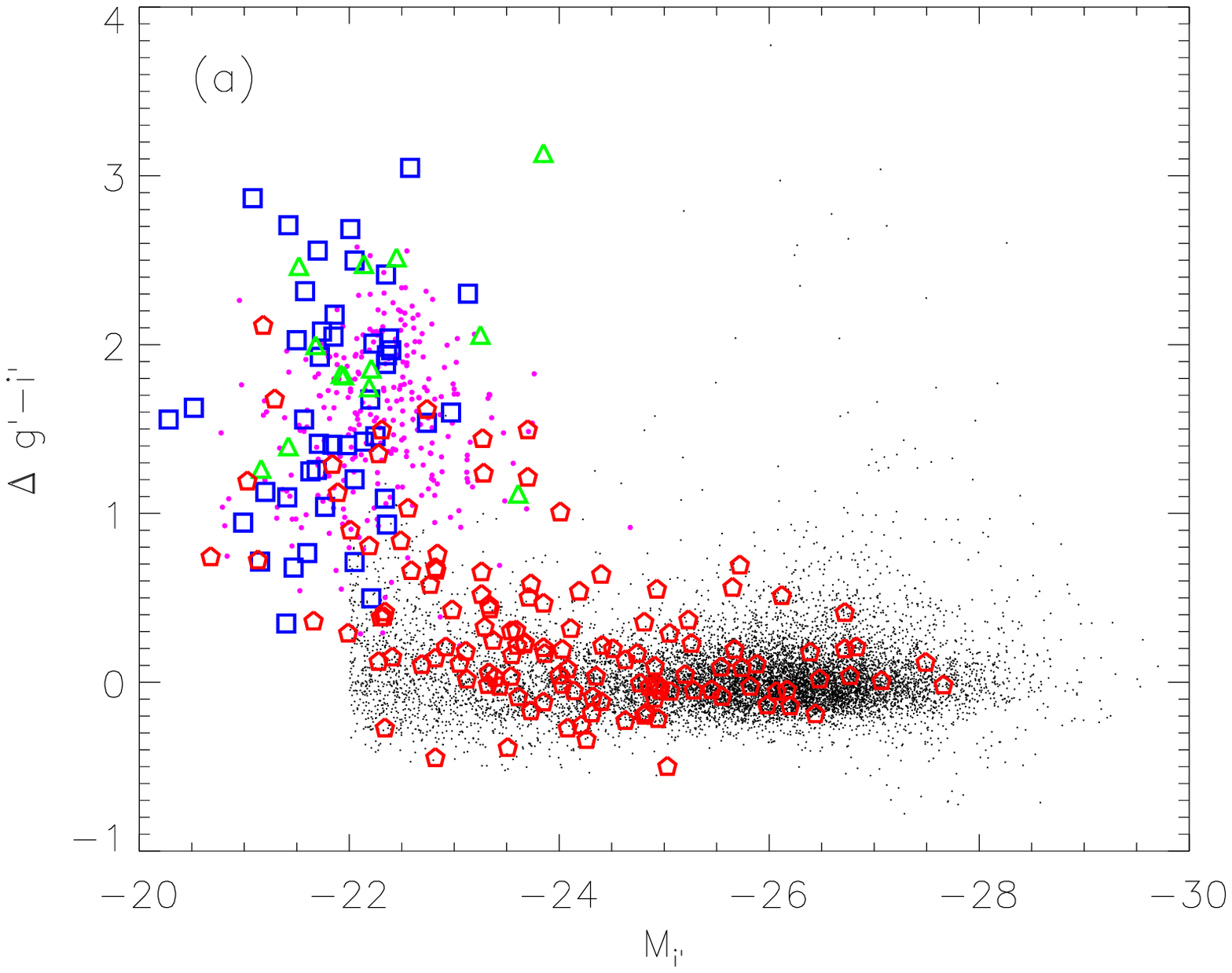}{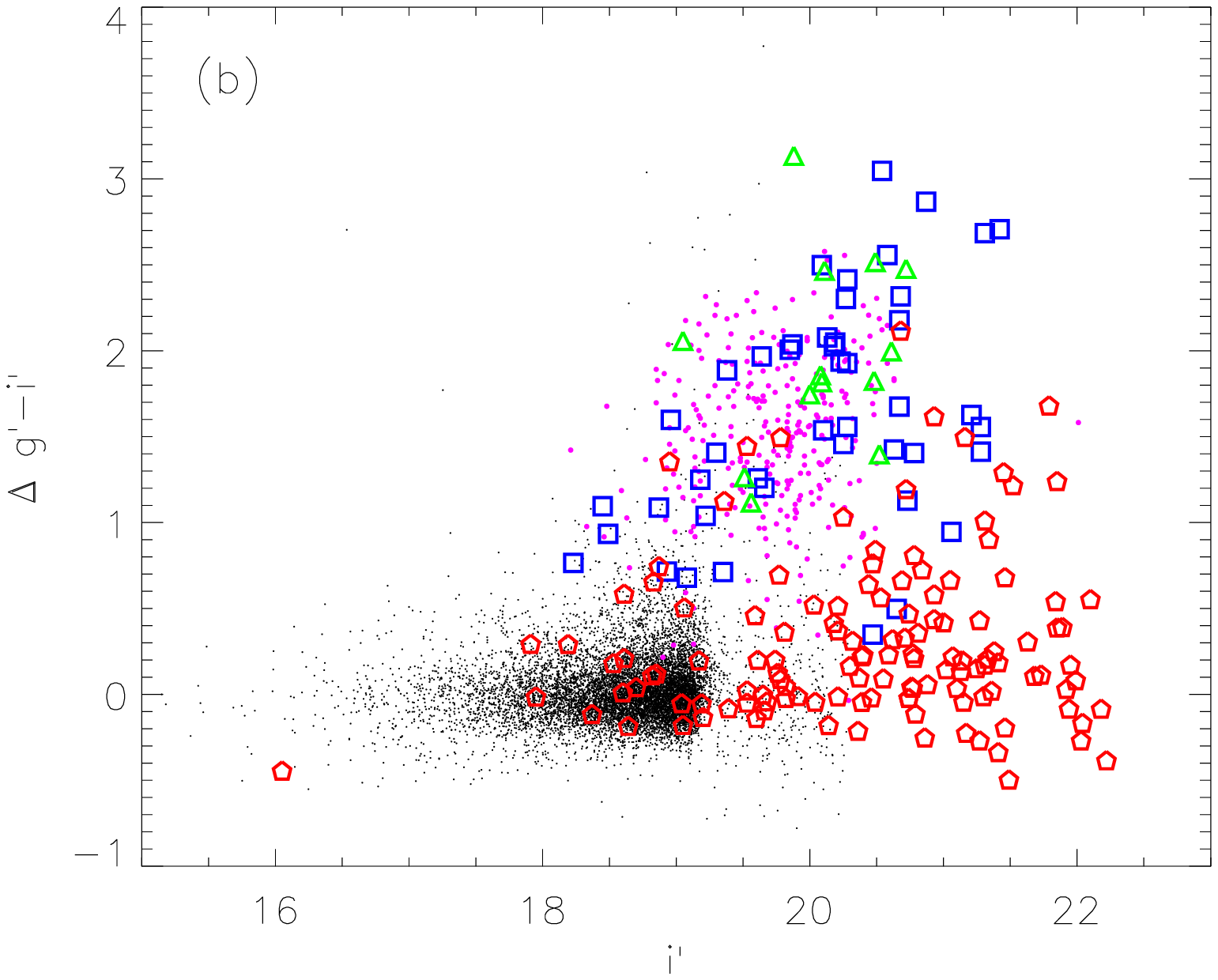}
\caption{Optical color offset as a function of absolute magnitude
M$_{i^{\prime}}$ (a) and apparent magnitude $i^{\prime}$ (b). Same
symbol types as Figure~\ref{sdss}a.}
\label{sdss2}
\end{figure}

\noindent We find that the distribution of ChaMP BLAGN has a red wing
(skewness=--1.7) which causes the mean to shift further to the red
than the SDSS quasars (skewness=7.1).  We calculate that 26\% of our
ChaMP AGN sample has $\Delta~g^{\prime}-i^{\prime}>0.2$ above that
expected from the SDSS distribution.  Therefore, an X-ray selected
sample can reveal a population of luminous, red AGN underrepresented
in optically selected samples.  At $z<1$ (Figure~\ref{sdss}), we see a
population of red BLAGN that would not be detected in the SDSS.  As
shown in Figure~\ref{sdss2}a, these red BLAGN tend to have low optical
luminosities ($M_{i^\prime}<-24$) and are probably missed in the SDSS due to
their continuum color, which can be influenced by their host galaxy.
In Figure~\ref{sdss2}b, we see that many of the red BLAGN are below
the flux limit of the SDSS.  A larger sample of optically bright ChaMP
AGN is required to definitively identify a population missed by the
optical surveys.

Most ChaMP NELG and ALG are red and fall off the color range for
optically selected type 1 quasars.  We would not expect these types of
AGN to be included in the SDSS quasar catalog due to their narrow or
nonexistent emission lines and most have $M_{i^\prime}>$--22.  These
objects appear to have colors dominated by their host galaxy.  To
illustrate, we plot in Figure~\ref{sdss}a the expected colors of an
elliptical (E0) and a spiral (Sb) galaxy using the Hyper\_z
(Bolzonella, Miralles, \& Pelló 2000) photometric code.  We have
assumed a Bruzual-Charlot spectral energy distribution, galactic
extinction ($E_{\rm{B-V}}$=0.05) using the Calzetti (1999) reddening
law and no intrinsic dust reddening.  Most NELG and ALG have colors
between these two tracks.  We also plot the type 2 SDSS QSO candidates
from Zakamska et al. (2003).  The absolute magnitude ($M_{i^\prime}$) is
calculated from the i$^{*}$ magnitude with a correction for Galactic
extinction ($E_{\rm{B-V}}$=0.05) and no $k$-correction.  The ChaMP
NELG and SDSS type 2 QSOs potentially represent the same population.
The extraordinary advantage of an X-ray selected survey is seen with
the detection of AGN with no optical emission lines at all (ALG).

\section{X-ray absorption and optical extinction}
\label{results}

We present evidence of a direct relationship between the absorption of
X-rays and optical extinction for the majority of the hard AGN, as
expected if unification models \citep{an93} extend to the X-ray
regime.  We attempt to reconcile the variety of AGN types from an
X-ray perspective.  We use the term absorbed for any AGN with X-ray
$N_{\rm{H}}>10^{22}$ cm$^{-2}$ and unabsorbed for those with columns
below this value.  As justified in Section~\ref{intrinnh}, this
$N_{\rm{H}}$ value represents the amount of absorbing neutral gas
needed to hide the broad emission line region (BLR), assuming a Milky
Way gas-to-dust ratio and $\alpha_{OX}=1.48$ for each AGN, in a
5L$_{\star}$ host galaxy.  As discussed below, most AGN with
unabsorbed X-ray emission show signs of optical emission from the BLR.
Alternatively, judging by the lack of broad emission lines, most X-ray
absorbed AGN have significant optical extinction.  However, there is a
small though significant fraction that do not fall into the simple
obscured or unobscured scenario.  The various classes illustrating the
permutations of X-ray absorption and optical extinction are delineated
in the following sections.

\subsection{Unabsorbed AGN with Broad Emission Lines}

\begin{figure}

\epsscale{0.9}

\plotone{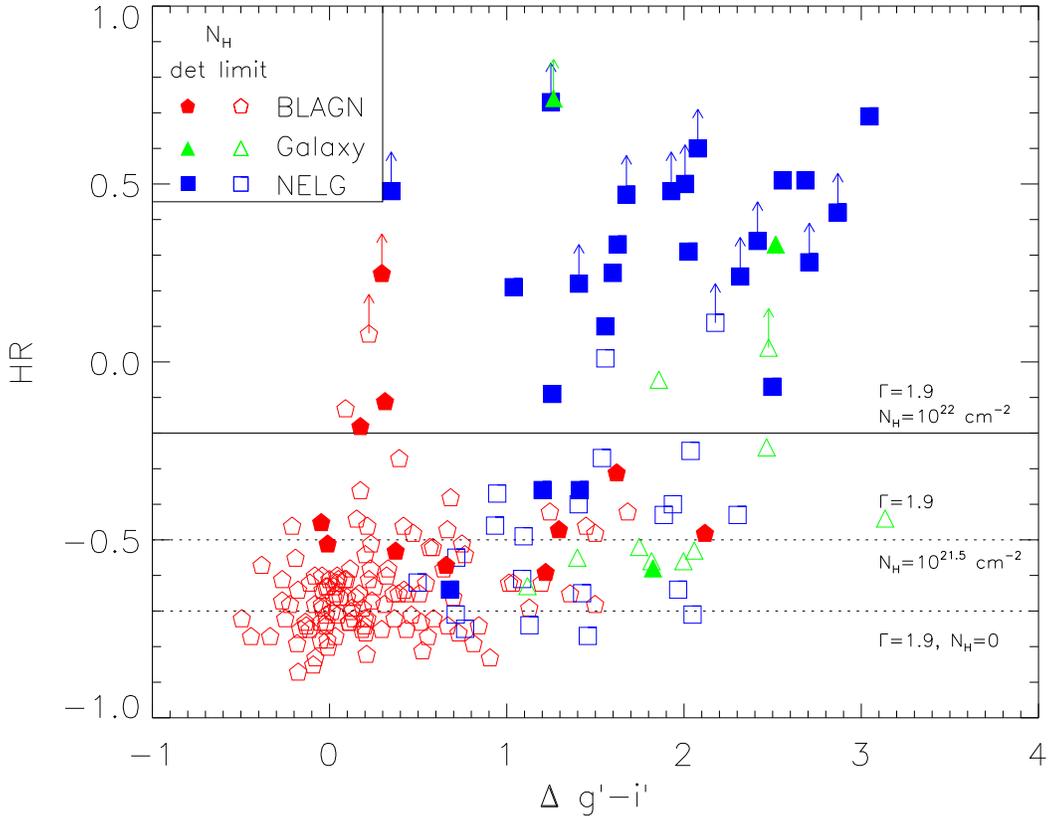}

\caption{Hardness ratio vs. optical color. The horizontal lines mark
the HR for a typical AGN with a powerlaw continuum ($\Gamma=1.9$)
absorbed by neutral gas intrinsic to the source.  The solid line marks
the division between unabsorbed and absorbed AGN as defined in this
paper.  The filled symbols mark the well constrained $N_{\rm{H}}$
measurements from X-ray spectral fits (Section~\ref{xfit}) to compare
with Figure~\ref{nh}.}
\label{hrcolor}
\end{figure}

We find that the majority of AGN with unabsorbed ($N_{\rm{H}}<10^{22}$
cm$^{-2}$) X-ray emission have no apparent extinction of their optical
light.  First, the optical spectra of 80\% of the unabsorbed AGN (HR
$<$ --0.2; Figure~\ref{hr}) are characterized by the presence of broad
emission lines (e.g., Ly$\alpha$, CIV, CIII], MgII).  In
Figure~\ref{hrcolor}, we plot the X-ray hardness ratio as a function
of optical color for our AGN sample.  These BLAGN have blue optical
colors and are grouped in a region (HR$<$--0.4 and
$g^{\prime}$--$i^{\prime}<1$) characteristic of unobscured AGN. As
mentioned previously, our BLAGN sample at $z>1$ has colors similar to
the optically-selected SDSS quasars (Figure~\ref{sdss}a) but with a
slightly broader and skewed distribution (Figure~\ref{color_comp}).
We conclude that no significant obscuring material is preventing us
from observing the broad line region in these unabsorbed AGN.
       
We find that 10\% of ChaMP BLAGN have red optical colors with
$g^{\prime}$--$i^{\prime}>1.0$ (Figure~\ref{hrcolor}), offset from the
locus of SDSS quasars (Figure~\ref{sdss}a).  We suspect that host
galaxy emission is contributing to their red optical color since these
ChaMP BLAGN, mostly at $z<1$, represent the lower luminosity fraction
of our sample ($L_{2-8 \rm{kev}}<10^{44}$ erg s$^{-1}$;
Figure~\ref{lzdistr}).  To further illustrate, Figure ~\ref{sdss}b
shows that the absolute magnitudes ($M_{i^{\prime}}$) and colors of these
red BLAGN are intermediate between the non-BLAGN and the more luminous
($M_{i^{\prime}}$) AGN.  We present in Figure~\ref{misclass}, optical
spectra of two such objects with composite (AGN+host) spectra.  Both
of these AGN have broad optical emission lines but their red optical
continuum is dominated by the host galaxy.  Extinction due to dust
might contribute to both their low luminosity and optical color.  Dust
extinction is probably responsible for the red colors of three
luminous ($L_{2-8 \rm{kev}}>10^{44}$ erg s$^{-1}$) BLAGN seen in
Figure~\ref{sdss}a with $1.5<z<2.0$.  The ChaMP BLAGN might be a
fainter version of the 2MASS selected, red and X-ray unabsorbed BLAGNs
\citep{wi02}.

\subsection{Unabsorbed AGN Lacking Broad Emission Lines}
\label{unabsorbed_nolines}

Surprisingly, a number of unabsorbed AGN show no broad optical
emission lines.  These objects are clearly seen in
Figure~\ref{hrcolor} with HR$<$--0.2 and
$g^{\prime}$--$i^{\prime}>$1.0.  The nature of these sources is the
least understood.  Their red optical colors are characteristic of
their host galaxy but the classification might be due to the fact that
at certain redshifts, we lose the AGN diagnostic emission lines
(e.g. H$\alpha$, Mg II) within the observed optical spectroscopic
window for sources with comparable host galaxy and AGN optical
emisssion as discussed in Section~\ref{dilution}.  Usually H$\beta$ is
the only line available, which can suffer from both extinction and
contamination due to the host galaxy.  We are pursuing spectroscopy
further to the red for some of these objects to detect H$\alpha$ for
emission line diagnostics to securely identify the ioinizing source as
either an AGN or starburst.  Many of these AGN are at $z>0.4$ with
H$\alpha$ at $\lambda>9200~{\rm \AA}$, making this type of analysis
difficult for optically faint AGN.

\begin{figure}

\epsscale{0.8}
\plottwo{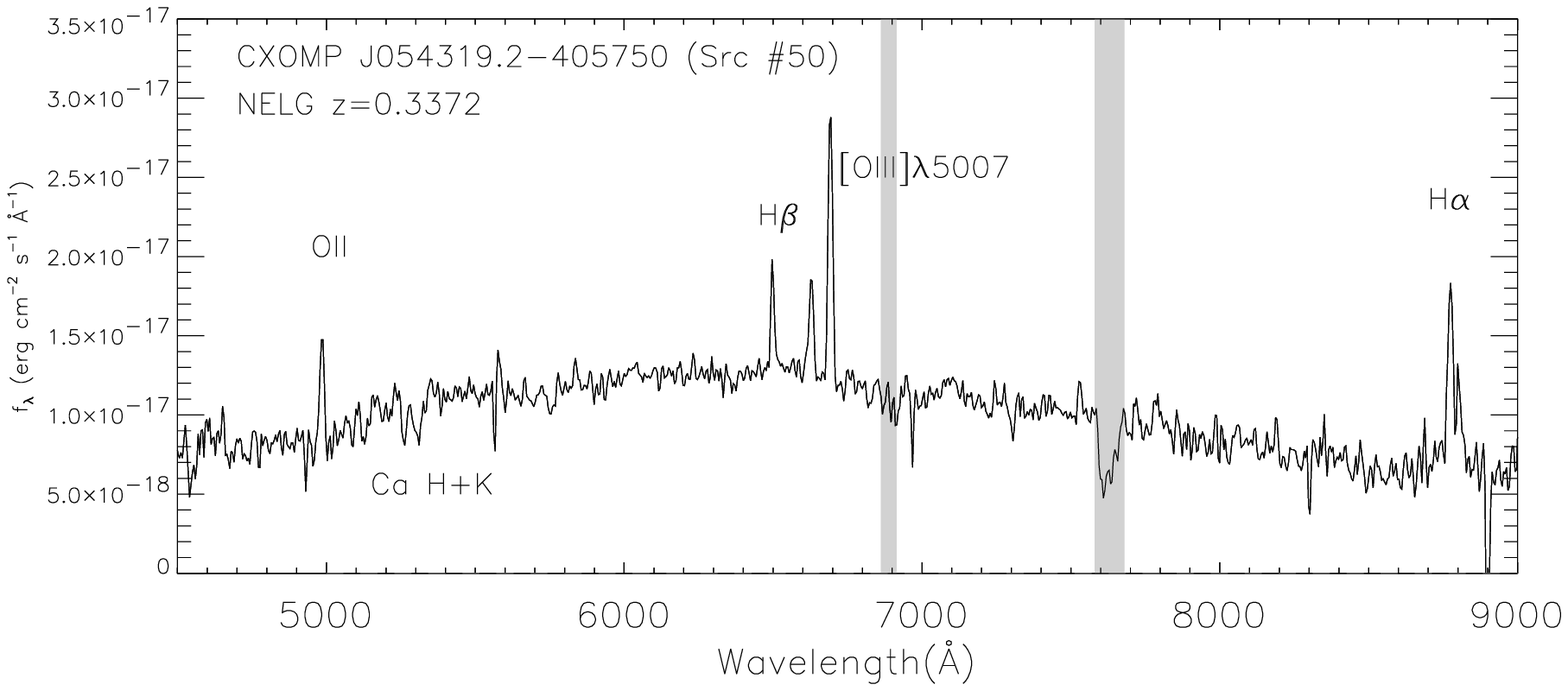}{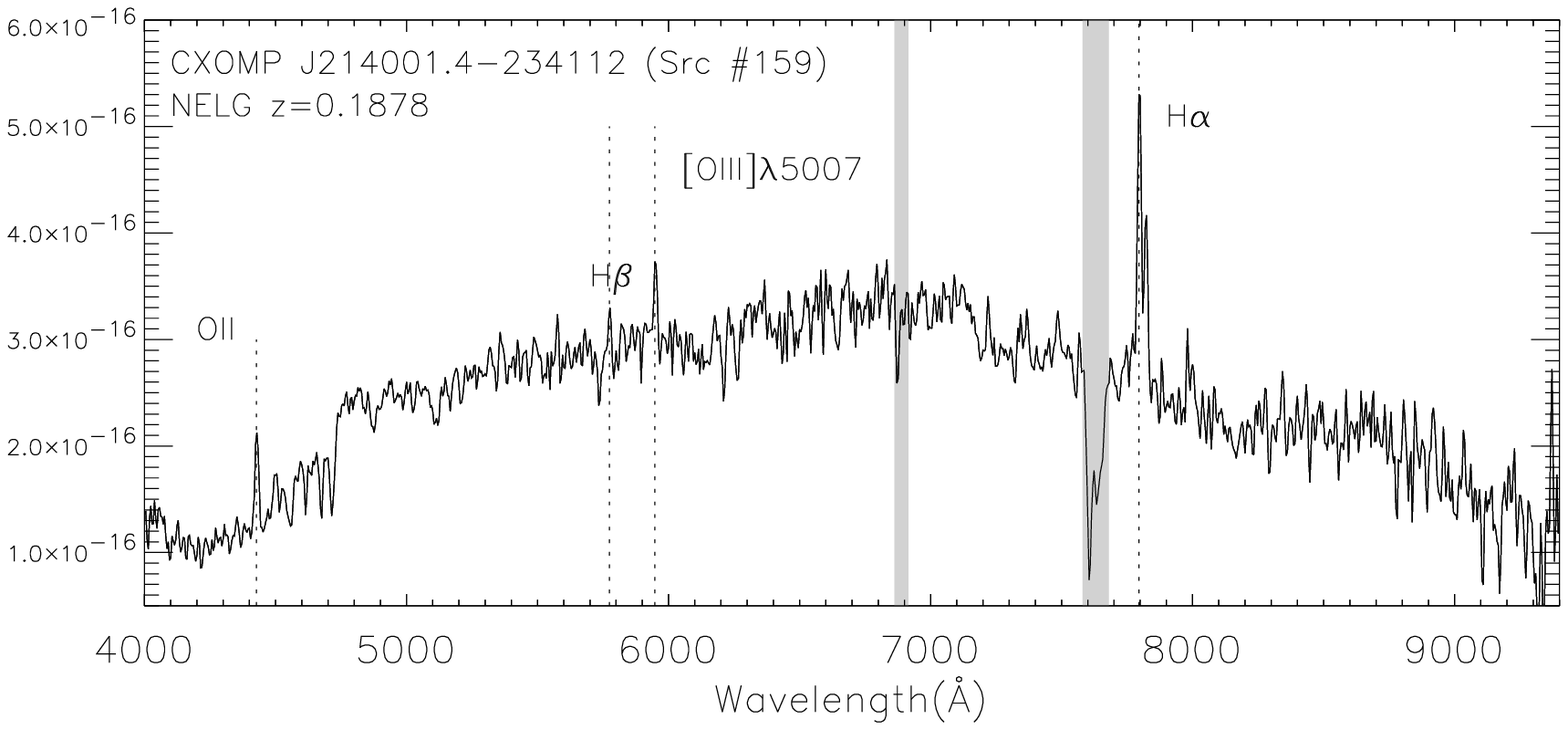}

\epsscale{0.9}
\plottwo{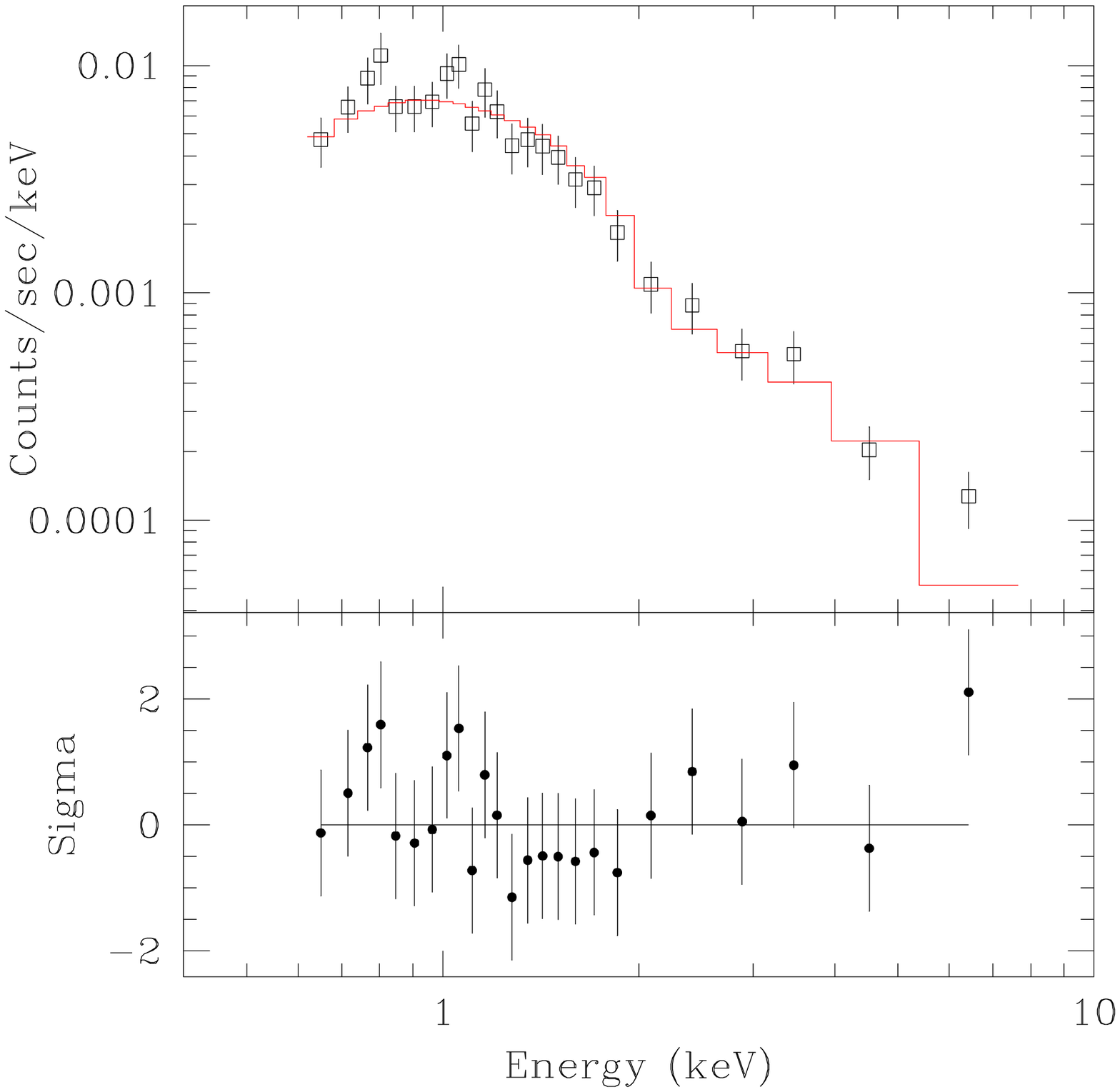}{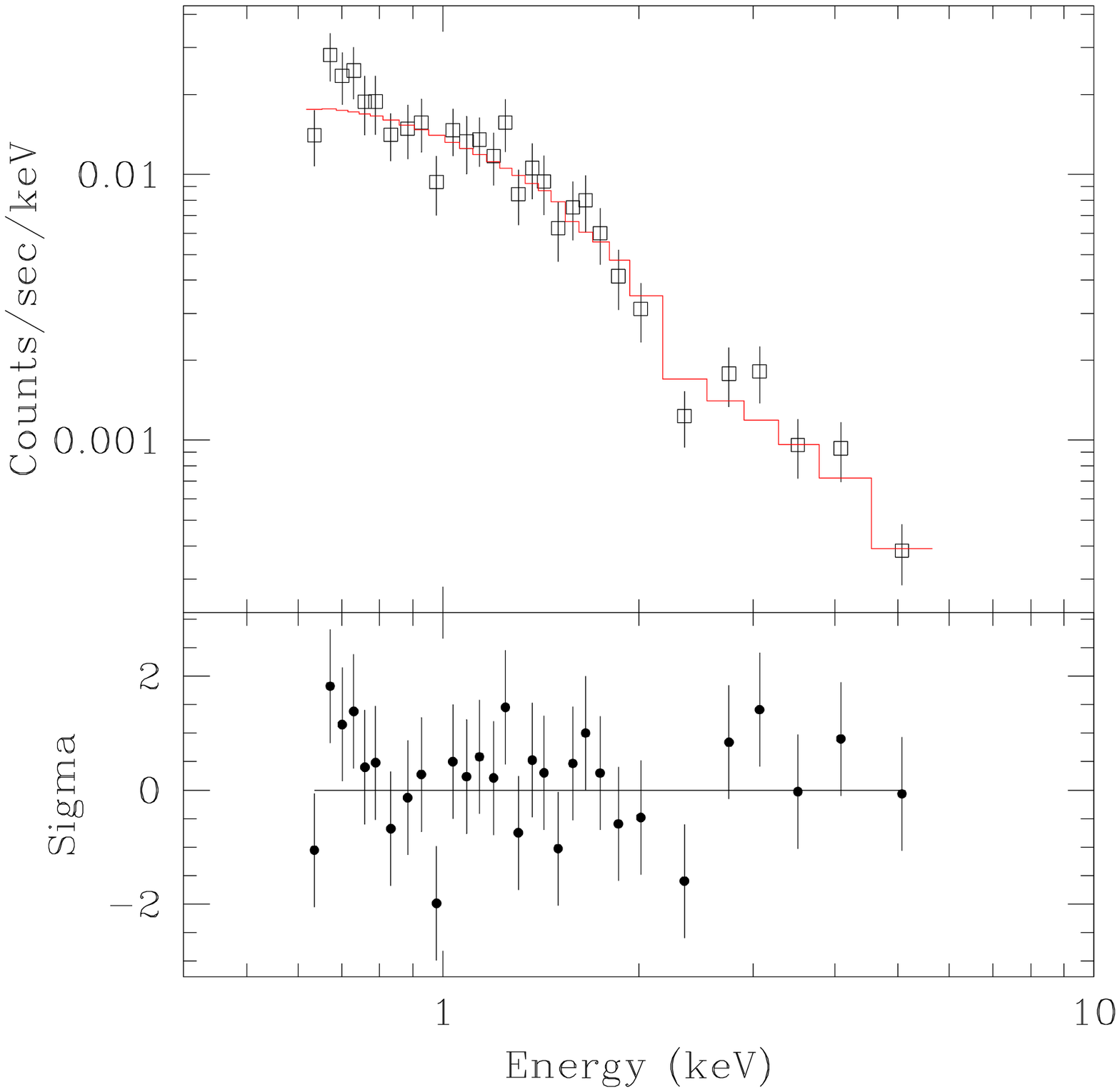}

\caption{The X-ray and optical spectra of two X-ray unabsorbed, Narrow
Emission Line Galaxies. Each column corresponds to one source.  (top)
Optical spectra, with shaded regions to mark the uncorrected
atmospheric absorption features, are displayed.  (Bottom) Best-fit
absorbed powerlaw model (top panel) and residuals (bottom panel) are
shown for each object.}
\label{noabsnelg}
\end{figure}

Two bright NELG may provide insight into the nature of this class of
unabsorbed AGN ($N_{\rm{H}}< 10^{22}$ cm$^{-2}$).  These objects might
be similar to the rare unabsorbed Seyfert 2 galaxies seen by other
groups in nearby objects (Georgantopoulos \& Zezas 2003; Panessa \&
Bassani 2002).  With over 400 counts in both sources [CXOMP
J054319.2-405750 (\#50; log L$_{\rm X}=43.05$;
$M_{i^{\prime}}=-22.05$), CXOMP J214001.4-234112 (\#159; log L$_{\rm
X}=43.08$; $M_{i^{\prime}}=-22.35$)], we are able to constrain their
X-ray spectral properties.  We fit the data using a spectral model
with the photon index and the intrinsic absorption as free parameters
(see Section~\ref{xfit}).  Figure~\ref{noabsnelg} shows the X-ray and
optical spectra.  Source \#50 is very soft with
$\Gamma=2.29^{+0.22}_{-0.16}$ and $N_{\rm{H}}< 8.2\times10^{20}$
cm$^{-2}$.  There could possibly be a soft excess below 2 keV.  Source
\#159 has a photon index $\Gamma=1.83^{+0.21}_{-0.13}$ with
$N_{\rm{H}}< 7.6\times10^{20}$ cm$^{-2}$.  These AGN show no signs of
X-ray absorption while their optical spectra show a strong galaxy
continuum with overlying narrow emission lines.  The width of H$\beta$
is 323 km s$^{-1}$ in source \#50 and the width of H$\alpha$ is 669 km
s$^{-1}$ in \#159.  Surprisingly, their low ionization emission line
ratios are not typical of an AGN, but resemble those of a star forming
galaxy.  We can explain the X-ray and optical properties with any of
the following scenarios: (1) severe dilution of the AGN emission by a
host undergoing prodigious star formation, (2) a Compton thick
obscuring medium in which we observe the indirect, reflected component
of the X-ray emission and the optical light is purely from the host
galaxy due to severe dust extinction, (3) a high dust to neutral gas
ratio which allows only the X-rays to penetrate or (4) beamed emission
from a BL Lac.  To properly discriminate between these models, a
sample of nearby AGN for which we can resolve the nuclear region and
properly subtract the host emission would be ideal.  For the Compton
thick case, the true L$_{\rm{X}}$ may be much higher.  In addition, we
may find a large L$_{\rm{IR}}$ from reprocessed emission.

We are finding a population of optically `dull' galaxies (Severgnini
et al. 2003; Silverman et al. 1998; Tananbaum et al. 1997; Elvis et
al. 1981), recently coined XBONGS \citep{co02}.  These are galaxies
with X-ray bright nuclei that have weak or nonexistent emission lines.
All but a couple of the 13 ALG identified in this study are not
heavily absorbed (Figure~\ref{hrcolor}). Similar X-ray properties have
been noted in recent studies of other examples (e.g., Severgnini et
al. 2003).  As shown in Figure~\ref{misclass}, we may be able to
detect AGN optical signatures in some of these ALG using expanded
wavelength coverage to include H$\alpha$, which is less susceptible to
dust extinction and stellar absorption features.  Using the definition
of Dressler \& Shechtman (1987), only two are possible BL Lac
candidates based upon their 4000 ${\rm \AA}$ break contrast.

\subsection{Absorbed AGN}

We find that most X-ray absorbed ($N_{\rm{H}}>10^{22}$ cm$^{-2}$) AGN
suffer from optical extinction.  Almost all of the AGN (84\%) with HR
$>$ --0.2, lack broad optical emission lines (Figure~\ref{hrcolor}).
These absorbed AGN tend to have narrow emission lines in their optical
spectra.  At these redshifts ($z>0.4$), the optical spectra rarely
include a detection of all the emission lines (H$\alpha$, NII, [OIII],
H$\beta$) needed to solidify the optical classification as an AGN as
opposed to a star forming galaxy.  Given their high X-ray luminosity
($L_{X}>10^{42}$ erg s$^{-1}$), we are confident that the X-ray
emission is due to supermassive black hole accretion.

In Figure~\ref{hrcolor}, we show that these absorbed AGN (HR $>$
--0.2) have red optical colors ($g^{\prime}$--$i^{\prime}>$ 1.5)
characteristic of their host galaxy and not the AGN itself
\citep{gr04,fi03}.  Most of these AGN have colors offset from the
locus of optically-selected QSOs and fall along the track of an early
type galaxy as shown in Figure~\ref{sdss}a.  These results agree with
studies which conclude that AGN are predominantly found in massive,
early type galaxies (Kauffmann et al. 2003).  For the most part, AGN
which suffer from absorption in the X-rays have significant optical
extinction that hides emission from the BLR.

\subsection{Intrinsic N$_{\rm{H}}$}

\label{intrinnh}

\begin{figure}
\epsscale{1.0}
\plotone{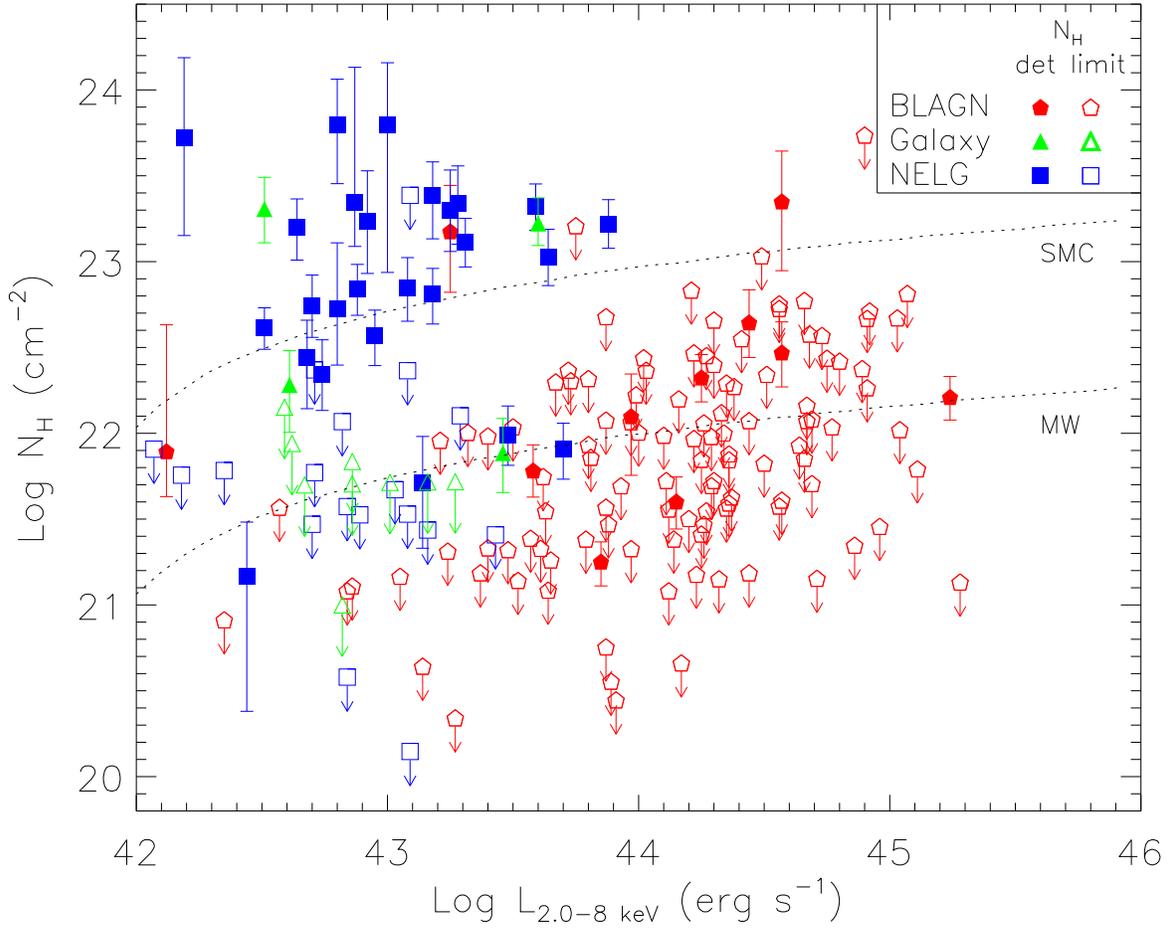}

\caption{Luminosity vs. $N_{\rm H}$. The X-ray luminosity has only
been corrected for galactic absorption.  Error bars represent 90\%
confidence intervals.  The filled symbols mark the well constrained
$N_{\rm{H}}$ measurements.  Upper limits are shown with hollow symbols
placed at the 90\% value.  The dashed lines mark our estimate of the
absorbing column needed to hide optical emission from the broad line
region for a given X-ray luminosity, assuming the average dust-to-gas
ratio from \citet{pe92} of the Milky Way (MW) and Small Magellanic
Cloud (SMC).}
\label{nh}
\end{figure}

We have measured the amount of intrinsic absorption towards each AGN
by fitting the X-ray count distribution with a fixed powerlaw model as
described in Section~\ref{xfit}.  In Figure~\ref{nh}, we plot the best
fit intrinsic $N_{\rm{H}}$ as a function of rest frame X-ray
luminosity.  Measurements with errors (1.6$\sigma$) are plotted as
filled symbols.  Upper limits are marked as hollow symbols placed at
the 1.6$\sigma$ upper limit with a downward arrow.

\begin{figure}

\epsscale{0.8}
\plotone{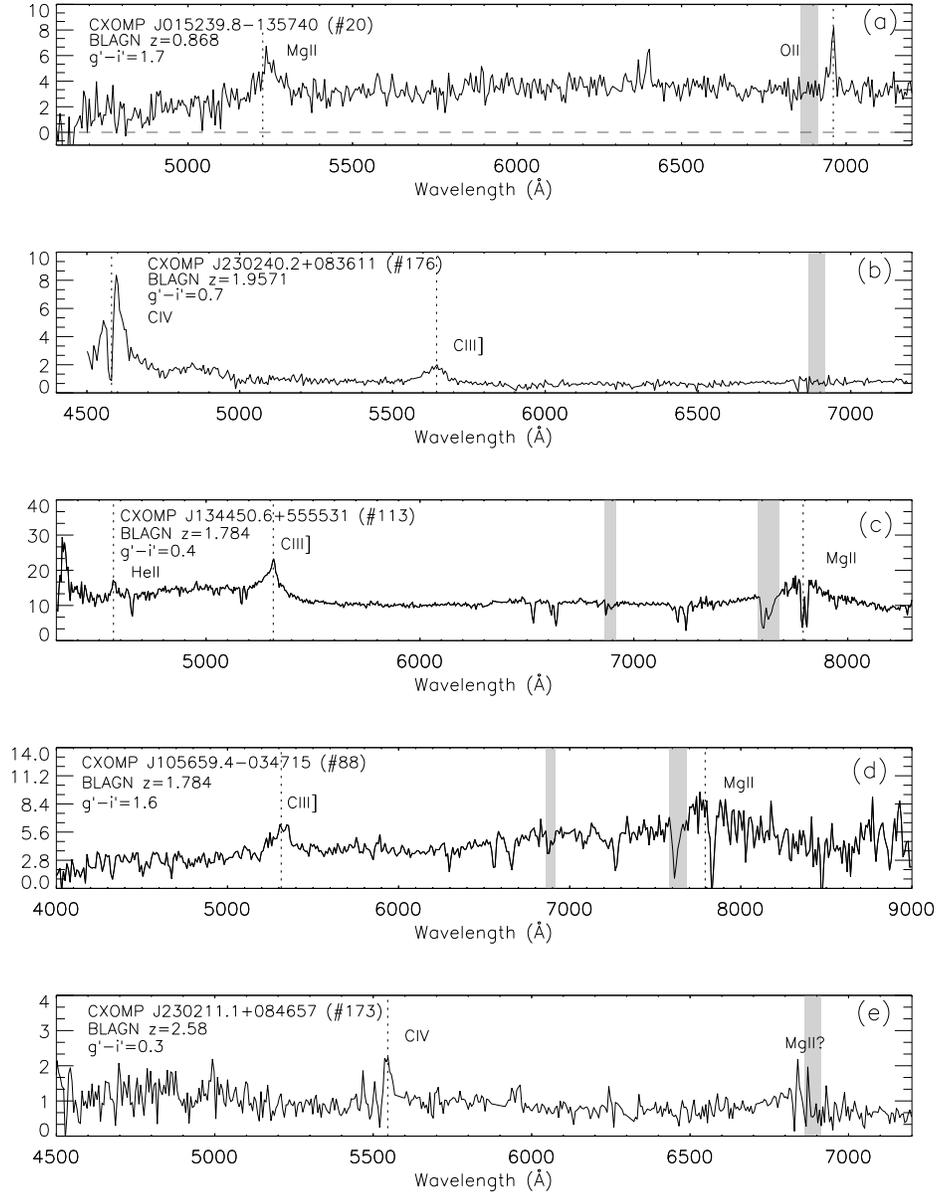}

\caption{Optical spectra of X-ray absorbed quasars with $N_{\rm H}> 10^{22}$ cm$^{-2}$ and rest frame L$_{2-8\rm{keV}}>10^{44}$ erg s$^{-1}$.  The flux (f$_{\lambda}$; y--axis) is in units of 10$^{-17}$ erg cm$^{-2}$ s$^{-1}$ $\rm{\AA}^{-1}$.  Dashed lines show the spectral features at the redshift of the object.  The optical color ($g^{\prime}-i^{\prime}$) is given.  Shaded regions mark the uncorrected telluric O$_{2}$ absorption bands.}
\label{type2}
\end{figure}

The absorbed AGN have columns in the range of
$10^{22}<N_{\rm{H}}<10^{24}$ cm$^{-2}$.  The majority of these are
optically classified as NELG.  We have uncovered 5 luminous AGN with
$L_{2-8\rm{keV}}>10^{44}$ erg s$^{-1}$ which are heavily absorbed with
well-constrained columns ($N_{\rm{H}}>10^{22}$ cm$^{-2}$).  These
quasars would not be singled out as having high absorption based on
their hardness ratios since they are mostly at $z>1.7$, where the
observed X-ray counts are less affected by absorption.  (In
Figure~\ref{hrcolor}, the BLAGN with well constrained values of
$N_{\rm{H}}$ (filled symbols) do not necessarily have HR $> -0.2$.)
While these quasars have broad emission lines in their optical spectra
(Figure~\ref{type2}), they all show signs of either absorption or
extinction in the optical.  Two sources (12a,d) have a reddened
optical continuum possibly attributed to dust extinction and similar
to the red quasar 3C~212 (Elvis et al. 1994). Their colors are
substantially offset from the SDSS quasar locus in
Figure~\ref{sdss}. Three (12b,c,d) have narrow absorption lines that
might be related to the X-ray absorption.  Source
CXOMP\,J230240.2+083611 (\#176 in 12b) has an absorption line
blue-shifted from the emission line center.  This is evidence for
ionized intrinsic outflows which are known to have high absorbing gas
columns ($N_{H}\sim6\times10^{22}$ cm$^{-2}$) in BALQSOs (Green et
al. 2001).  Even in non-BALQSOs, \citep{laor02} found an association
in quasars between (soft) X-ray weakness and CIV absorption equivalent
width.  While the spectrum of source CXOMP\,J230211.1+084657 (\#173,
Figure~12e) has a poor S/N ratio, absorption might be seen blueward of
the CIV emission line.  The object CXOMP\,J134450.6+555531 (\#113 in
12c) has narrow absorption lines detached from the emission
lines. Either these absorption lines are extremely fast outflows
intrinsic to the AGN or a more probable explanation is an intervening
absorber.  Based on these 5 absorbed quasars, we find that 6\% (5/79)
of luminous ($L_{2-8\rm{kev}}>10^{44}$ erg s$^{-1}$) X-ray-selected
BLAGN have $N_{\rm{H}}>10^{22}$ cm$^{-2}$ (2\% of all AGNs).  This is
consistent with the value of 10\% found by \citet{pe04} and
\citet{pa03}.

As expected, the unabsorbed (N$_{\rm{H}}<10^{22}$ cm$^{-2}$;
Figure~\ref{nh}) AGN population is dominated by BLAGN.  Almost all of
the measurements are poorly constrained (upper limits) with a low
probability of significant absorption.  The apparent trend of
increasing $N_{\rm{H}}$ with luminosity for the BLAGN is probably due
to our decreasing sensitivity to the absorbing column at higher
redshifts, rest frame soft X-rays, most susceptible to absorption, are
redshifted out of the $Chandra$ bandpass.  There are a fair number of
NELG and ALG with no measurable absorbing column.  As described in the
previous section, we suspect that host dilution is confusing the
spectroscopic characterization of these sources.

We crudely estimate the $N_{\rm{H}}$ needed to hide the optical
emission from an embedded AGN of a given X-ray luminosity.  From the
rest frame $L_{2.0-8.0 \rm{keV}}$ and $\alpha_{OX}$=1.48 (the mean
X-ray to optical flux ratio for the X-ray selected BLAGN;
Section~\ref{dilution}), we calculate the expected unabsorbed
$l_{\nu}$ (2500 $\rm{\AA}$).  We then compare the 2500 $\rm{\AA}$
monochromatic luminosity to that of a host galaxy, assuming
$L_{\rm{Host}}=5L_{\star}$.  This luminosity corresponds to the
average host absolute magnitude ($M_{B}=-23.0$) found by \citet{ja03}
for a sample of bright QSOs.  Using the gas-to-dust ratio of the Milky
Way ($N_{\rm{H}}=4.82\times10^{21}$ $E_{\rm{B-V}}$ mag$^{-1}$
cm$^{-2}$) and Small Magellanic Cloud ($N_{\rm{H}}=4.52\times10^{22}$
$E_{\rm{B-V}}$ mag$^{-1}$ cm$^{-2}$) from \citet{pe92}, we determine
the amount of dust extinction and hence the column density required to
diminish the AGN optical light to a tenth of the assumed host optical
luminosity.  The dashed lines in Figure~\ref{nh} shows the results
from this rather simplified calculation for both gas-to-dust ratios.
Based on the gas-to-dust ratio of the Milky Way, the column densities
needed to hide an AGN are $N_{\rm{H}}\sim10^{22}$ cm$^{-2}$ which
marks our division between absorbed and unabsorbed AGN.  The column
densities could potentially be higher (Fig~\ref{nh}; top, dashed
line), given that luminous quasars found in the SDSS have dust
extinction more typical of the Small Magellanic Cloud \citep{ho04}.

\subsection{Optically faint X-ray sources and type 2 quasars}
\label{qso2}

As evident from Figure~\ref{fxopt}, there are many unidentified X-ray
sources with $r^{\prime}>$ 22.  We believe that a significant fraction
of these sources may be the elusive type 2 quasars.

A number of ongoing X-ray surveys (e.g., CDF-N and CDF-S) have
acquired spectroscopic redshifts fainter than the current ChaMP sample
and have identified a few as luminous narrow line AGN.  The HELLAS2XMM
(Fiore et al. 2003) survey has spectroscopically identified 8 type 2
QSOs from a sample of 13 over 0.9 deg$^{2}$ with
$f_{\rm{X}}/f_{r^{\prime}} >$ 10.  The redshift range of these quasars
($0.7<z<1.8$) as described previously has been inaccessible to ChaMP
(Figure~\ref{lzdistr}) due to our bright spectroscopic limit.

\begin{figure}
\epsscale{1.0}
\plotone{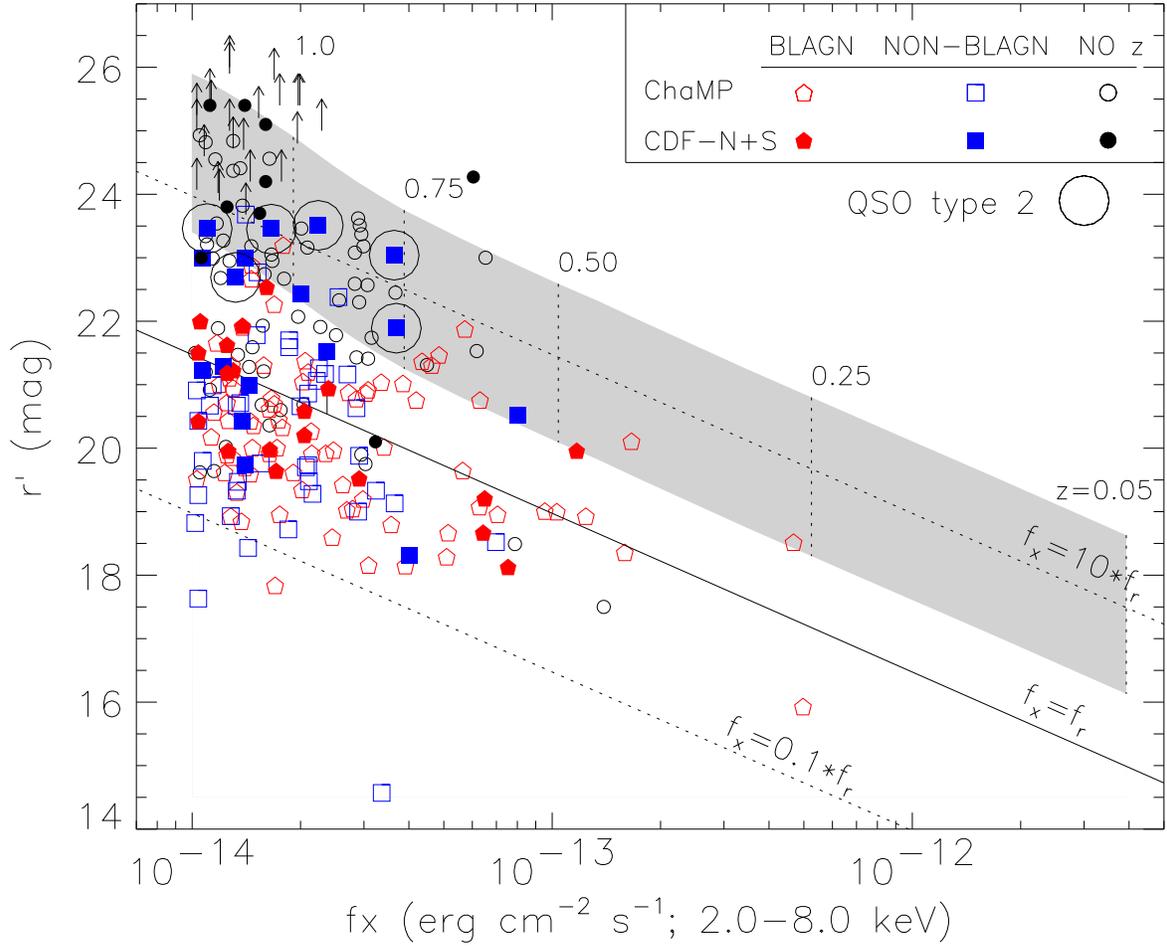}

\caption{Where are the type 2 QSOs?  The shaded region marks the
location of a hypothetical, obscured QSO out to $z=1.45$ as described
in Section~\ref{qso2}.  We have included the sources from the
$Chandra$ Deep Fields which include 6 such quasars.}

\label{predict}
\end{figure}

To investigate the lack of obscured, highly luminous quasars in our
sample, we illustrate their expected location in $f_{X}$--$r^{\prime}$
space (Figure~\ref{predict}).  We plot only the X-ray bright
($f_{2-8\rm{keV}}>1\times10^{-14}$ erg cm$^{-2}$ s$^{-1}$)
extragalactic sources (with the exclusion of clusters) and the
unidentified objects.  We have shaded the region for a hypothetical
quasar with $L_{\rm{X}}=10^{44}$ erg s$^{-1}$ and optical emission
purely from a host elliptical galaxy ($L_{\star}<
L_{\rm{r}}<10L_{\star}$) using Hyper\_z (Bolzonella, Miralles, \&
Pelló 2000).  We have labelled the redshifts ($0.05<z<1.0$) in this
region with dashed vertical lines.  We infer that most obscured
quasars probably fall within the region with $r^{\prime}>22$ based on
our X-ray sensitivity and area coverage.

To test this hypothesis, we have included data from the CDF-N and
CDF-S which have detected 6 type 2 quasars with
$f_{x}>1\times10^{-14}$ erg cm$^{-2}$ s$^{-1}$, $L_{2-8
\rm{kev}}>10^{44}$ erg s$^{-1}$ and no evidence for any broad optical
emission lines.  The optical magnitudes of these sources have been
converted to the SDSS photometric system using the transformations in
Fukugita (1996).  For the CDF-S data, we use average colors for each
object class from the CDF-N sources to convert the R band magnitude
(Szokoly et al. 2003).  It is apparent from the six possible type 2
quasars in the deep fields that there is potentially a significant
number of obscured quasars yet to be identified at faint optical
magnitudes.

The optically undetected, X-ray bright objects must await infrared
followup to learn more about their source properties.  However, we can
look at the X-ray hardness ratios to infer their host optical
properties. From Figure~\ref{hr}, we see that the optically undetected
sources have a wide range of hardness ratio with many extreme cases
(HR $>$ 0).  Since these X-ray sources have a HR distribution most
similar to the NELG at $f_{x}>1\times10^{-14}$ erg cm$^{-2}$ s$^{-1}$,
it seems likely that many of these optically undetected sources are
the narrow emission line objects at higher redshift possibly
comprising a significant fraction of the highly luminous and obscured
quasar population.

\section{Conclusions}

We have presented an analysis of the X-ray and optical properties of
188 hard X-ray selected AGN in 20 medium depth ChaMP fields.  These
AGN have been classified by optical spectroscopy to be comprised of
BLAGN (69\%), NELG (24\%) or ALG (7\%).  Overall, we find that 81\% of
the AGN agree with simple AGN unification models \citep{an93}.
The two main points of evidence are as follows:

\begin{itemize}

\item 80\% of unabsorbed AGN ($N_{\rm{H}}<10^{22}$ cm$^{-2}$) have
optical properties characterized by broad emission lines.  The optical
colors of ChaMP BLAGN are predominately blue
($g^{\prime}-i^{\prime}<1.0$) and similar to optically-selected
quasars from the SDSS but with a slighly wider distribution.  A
significant red ($g^{\prime}$--$i^{\prime}>1.0$) BLAGN population
exists with optical properties influenced by their host galaxy.  These
objects represent 10\% of the full AGN sample.

\item 84\% of absorbed AGN ($N_{\rm{H}}>10^{22}$ cm$^{-2}$) lack broad
optical emission lines.  Most are associated with NELG with column
densities in the range 10$^{22}<N_{\rm{H}}<10^{24}$ cm$^{-2}$.  Their
optical colors ($g^{\prime}-i^{\prime}>1.0$) are characteristic of a
luminous, early type galaxy.

\end{itemize}

We also find a number of atypical AGN (19\%) whose X-ray and optical
properties can be explained without any need to alter the AGN
unification models.

\begin{itemize}

\item The lack of broad optical emission lines in X-ray unabsorbed,
NELG and ALG can be attributed to a strong host galaxy contribution in
17\% of the total AGN sample.

\item The large amount of X-ray absorption ($N_{\rm{H}}>10^{22}$
cm$^{-2}$) in a few BLAGN (2\% of all AGNs) may be related to
signatures of absorption in their optical spectra.  A future ChaMP
study will determine if the X-ray absorption is associated with warm
(ionized) intrinsic outflowing gas similar to that seen in Broad
Absorption Line quasars.

\end{itemize}

\acknowledgments

We are greatly indebted to NOAO and the SAO TAC for their support of
this work.  Many thanks to Robert Kirshner, Warren Brown, John Huchra,
Kevin Krisciunas, and Guillermo Torres for contributing to the ChaMP
spectroscopic program.  We remain indebted to the staffs at Kitt Peak,
CTIO, Las Campanas, Keck, FLWO, and MMT for assistance with optical
observations.

We gratefully acknowledge support for this project under NASA under
CXC archival research grants AR3-4018X and AR4-5017X.  TLA, WAB, RAC,
PJG, DWK, AEM, HT, and BW also acknowledge support through NASA
Contract NASA contract NAS8-39073 (CXC).  DW receives support through
the SAO Research Experiences for Undergraduates (REU) Summer Intern
Program that is supported by the NSF.

\clearpage

\begin{deluxetable}{llcllllc}
\tabletypesize{\scriptsize}
\tablecaption{\Chandra\, Fields \label{xfields}}
\tablewidth{0pt}
\tablehead{
\colhead{ObsID} & 
\colhead{PI Target}  &
\colhead{~~ Exposure\tablenotemark{a}} & 
\colhead{ACIS CCDs\tablenotemark{b}} & 
\colhead{RA}  & 
\colhead{DEC} & 
\colhead{UT Date} &
\colhead{Galactic N$_{\rm H}$}\tablenotemark{d}\\
\colhead{}  &
\colhead{}  &
\colhead{(ksec)}  &
\colhead{}  &
\multicolumn{2}{c}{ J2000\tablenotemark{c} } & 
\colhead{}  &
\colhead{($10^{20}$cm$^{-2}$)}\\
}
\startdata
520  &	 MS0015.9+1609&61.0&012{\underline3}&00:18:33.4&+16:26:34.8&2000 Aug 18&4.06\\
913  &   CLJ0152.7-1357&34.6&012{\underline3}67&01:52:43.0&$-$13:57:30.0&2000 Sep 08&1.61\\
796  &   SBS 0335$-$052&47.0&012{\underline3}&03:37:44.0&$-$05:02:39.0&2000 Sep 07&4.98\\
624  &   LP944$-$20     & 40.9  & 236{\underline 7}  & 03:39:34.7&$-$35:25:50.0 & 1999 Dec 15 & 1.44\\
902  &   MS0451.6$-$0305&41.5&236{\underline7}&04:54:10.9&$-$03:01:07.2&2000 Oct 08&5.18\\
914  &   CLJ0542$-$4100 & 48.7  & 012{\underline 3}  & 05:42:50.2 &$-$41:00:06.9 & 2000 Jul 26 & 3.59\\
1602 &   Q0615+820&43.1&236{\underline 7}&06:26:02.9&+82:02:25.5&2001 Oct 18&5.27\\
377  &   B2 0738+313&26.9&36{\underline7}&07:41:10.7&+31:12:00.4&2000 Oct 10&4.18\\
2130 &   3C207&30.0&2{\underline3}67&08:40:48.0&+13:12:23.0&2000 Nov 04&4.14\\
512  &   EMSS1054.5-0321&75.6&236{\underline 7}&10:57:00&$-$03:37:00.0&2000 Apr 21&3.67\\
536  &   MS1137.5+6625  & 114.6 & 012{\underline 3} & 11:40:23.3 &+66:08:42.0 & 1999 Sep 30 & 1.18\\
809  &   MRK 273X&40.9&236{\underline 7}&13:44:47.5&+55:54:10.0&2000 Apr 19&1.09\\
541  &   V1416$+$4446   & 29.8  & 012{\underline 3}  & 14:16:28.8 &+44:46:40.8 & 1999 Dec 02 & 1.24\\
548  &   RXJ1716.9+6708&50.3&012{\underline3}&17:16:52.3&+67:08:31.2& 2000 Feb 27&3.71\\
830  &   Jet of 3C390.3&23.6&236{\underline7}&18:41:48.1&+79:47:43.0&2000 Apr 17&4.16\\
551  &   MS2053.7$-$0449&42.3&012{\underline3}6&20:56:22.2&$-$04:37:44.4&2000 May 13&4.96\\
928  &   MS2137$-$2340  & 29.1 &  236{\underline 7}  & 21:40:12.7 &$-$23:39:27.0 & 1999 Nov 18 & 3.57\\
431  &   Einstein Cross&21.9&236{\underline7}&22:40:30.4&+03:21:31.0&2000 Sep 06&5.34\\
918  &   CLJ2302.8+0844&106.1&012{\underline3}&23:02:48.1&+08:44:00.0&2000 Aug 05&5.50\\
861  &   Q2345$+$007    & 65.0  & 26{\underline 7}  & 23:48:19.6 &+00:57:21.1 & 2000 Jun 27 & 3.81\\
\enddata
\tablenotetext{a}{Effective screened exposure time for the on-axis chip.}
\tablenotetext{b}{The ACIS CCD chips used in the observation with the aimpoint chip underlined. CCD 8 has been excluded (see text).}
\tablenotetext{c}{Nominal target position, not including any \Chandra\, pointing offsets.}
\tablenotetext{d}{Galactic column density taken from \citet{di90}.}
\end{deluxetable}

\clearpage

\begin{deluxetable}{lccccccccc}
\tabletypesize{\scriptsize}
\tablecaption{KPNO 4m Optical Imaging \tablenotemark{a}\label{timages}}
\tablewidth{0pt}
\tablehead{
\colhead{ObsID} & 
\colhead{UT Date}  &
\colhead{Filter}  &
\colhead{Dithers} &
\colhead{~ ~ Exposure} & 
\colhead{Airmass}  & 
\colhead{FWHM\tablenotemark{b}}  & 
\colhead{m$_{\rm TO}$\tablenotemark{c}} &
\colhead{m$_{5\sigma}$\tablenotemark{d}} \\
\colhead{}  &
\colhead{}  &
\colhead{}  &
\colhead{}  &
\colhead{(total sec)}  &
\colhead{(Mean)}  &
\colhead{(\arcsec)}  &
\colhead{Limit}  &
\colhead{Limit}  &
\colhead{}  \\
}
\startdata
377&  21 Feb 2001          &\gp&3&1800&1.00&1.1&25.4&26.5\\
         &             & \rp &3&1500&1.01&1.0&24.9&26.0\\
         &             & \ip &3&1500&1.04&1.1&24.1&25.2\\
431  & 11 Jun 2000&\gp &2 & 1000&1.36&1.6& 24.1 &25.4\\
         &             & \rp &1  &500&1.28&1.6& 23.4 &24.7  \\
         &             & \ip & 1&360&1.25&1.2&22.9 & 24.1 \\
512&  21-22 Feb 2001          &\gp&5&4500&1.29&1.3&24.9&26.3\\
         &             & \rp &3&2400&1.23&1.1&24.4&25.8\\
         &             & \ip &5&2000&1.29&1.3&23.6&24.8\\
520& 25 Oct 2001       &\gp&3&1950&1.24&1.2&24.1&25.2\\
         &             & \rp &3&1200&1.34&1.2&23.9&24.9\\
         &             & \ip &3&900&1.46&1.4&22.6&24.0\\
548  & 10 Jun 2000&\gp & 2&1800 &1.28&1.2&24.8  &26.2\\
         &             & \rp &2  &1200&1.24&1.3& 24.1 & 25.4 \\
         &             & \ip &2 &1200&1.22&1.7&23.4 & 24.6 \\
551  & 10,12 Jun 2000&\gp & 3& 2100&1.28&1.7& 24.4 &25.6\\
         &             & \rp & 4 &1200&1.25&1.8&23.9  & 25.0 \\
         &             & \ip & 2&600&1.24&1.4&22.9 & 24.5 \\
796  & 24 Oct 2001&\gp &3&2700 &1.33&1.1&25.1  &26.4\\
         &             & \rp &3&2400&1.52&1.2&24.6  &25.7  \\
         &             & \ip &3&1200&1.28&1.1&23.6 & 24.8 \\
809&  11 June 2000         &\gp&3&2100&1.19&1.8&23.1&24.4\\
         &             & \rp &3&1500&1.14&1.7&23.4&24.6\\
         &             & \ip &3&1500&1.10&1.8&23.1&24.4\\
830&  11 June 2000          &\gp&3&1800&1.49&1.6&24.1&25.6\\
         &             & \rp &2&1000&1.49&1.6&23.4&24.8\\
         &             & \ip &1&600&1.49&1.6&22.9&24.2\\
902  & 23 Oct 201&\gp & 3 &1800&1.24&1.0& 24.9 &26.2\\
         &             & \rp &3  &1800&1.22&1.1&24.4  & 25.6 \\
         &             & \ip &3 &1200&1.24&1.0& 23.6& 24.9 \\
913  &23 Oct 2001& \gp &5&3500 &1.46&1.1&25.1  & 26.4\\
         &             & \rp &5&3000&1.44&1.0&24.6  &  25.9\\
         &             & \ip &5&2000&1.52&1.1&23.6 &  24.8\\
918  & 23 Oct 2001&\gp & 5&4500&1.20&1.2& 24.9 &26.0\\
         &             & \rp & 5 &3000&1.09&1.3&24.1  & 25.6 \\
         &             & \ip & 5&1500&1.12&1.2&23.1 & 24.8 \\
1602  & 14 Dec 2001&\gp &1 & 520&1.56&2.4& 23.3 &24.8\\
         &             & \rp &3  &1800&1.56&2.1&23.1  &24.8  \\
         &             & \ip &3 &1800&1.56&2.5&22.6 & 24.7 \\
2130& 22 Feb 2001           &\gp&3&3000&1.16&1.4&25.1&26.5\\
         &             & \rp &3&2700&1.08&1.3&24.9&25.9\\
         &             & \ip &5&2500&1.06&0.9&24.6&25.8\\

\enddata
\tablenotetext{a}{We tabulated 14 additional ChaMP fields following Green et
al. (2004).}
\tablenotetext{b}{FWHM of point sources in final stacked images.}
\tablenotetext{c}{Turnover magnitude limit at $\sim90\%$ completeness,
using 0.25\,mag bins before extinction correction, as described in the text.} 
\tablenotetext{d}{Magnitude limit for a $\sim5\,\sigma$ detection.} 
\end{deluxetable}

\begin{deluxetable}{llllccc}
\tabletypesize{\scriptsize}
\tablecaption{Optical Spectroscopic Followup \label{ospec}}
\tablewidth{0pt}
\tablehead{
\colhead{Telescope} &\colhead{Instrument} &
\colhead{Grating/Grism}&
\colhead{$\lambda$ range} &\colhead{R} &\colhead{Spectral} &
Number\\
\colhead{} &\colhead{} &\colhead{} &\colhead{(\AA)}
&\colhead{($\lambda/\Delta\lambda$)} &\colhead{Resolution (\AA)}&
of spectra\\
}
\startdata
WIYN\tablenotemark{1}&HYDRA&316@7.0&4500-9000&950&7.8&91\\
CTIO/4m&HYDRA&KPGL3&4600-7400&1300&4.6&48\\
Magellan&LDSS-2&med red \& blue&3600-8500\tablenotemark{2}&520&13.5&41\\
Magellan&B\&C&300~l/mm&3700-8700&384&13.0&7\\
MMT&Blue Channnel&300~l/mm&3500-8300&800&8.8&15\\
Keck I&LRIS&300/5000&4000-9000&484&13&5\\
FLWO 1.5m&FAST&300~l/mm&3600-7500&850&5.9&2\\
\enddata
\tablenotetext{1}{The WIYN Observatory is a joint facility of the
University of Wisconsin Madison, Indiana University, Yale University,
and the National Optical Astronomy Observatory.}
\tablenotetext{2}{Spectral coverage can vary as a function of slit
position in the mask.}
\end{deluxetable}

\begin{deluxetable}{lccc}
\tabletypesize{\scriptsize}
\tablecaption{X-ray source populations \label{numbers}}
\tablewidth{0pt}
\tablehead{
\colhead{Class} &
\colhead{All\tablenotemark{a}}  &
\colhead{$r^{\prime}<22.5$}  &
\colhead{$r^{\prime}<22.5$}\\
&&&\colhead{L$_{\rm x}> 10^{42}$ erg s$^{-1}$}}
\startdata
BLAGN&139&132&130\\
NELG&53&52&45\\
ALG&18&16&13\\
STAR&10&10&$\ldots$\\
Cluster&5&5&$\ldots$\\
No z\tablenotemark{b}&195&63&$\ldots$\\
No opt\tablenotemark{c}&82&$\ldots$&$\ldots$\\
Total&497\tablenotemark{d}&273&188\\
\enddata
\tablenotetext{a}{f$_{2.0-8.0 {\rm keV}}>2.7\times10^{-15}$ erg cm$^{-2}$ s$^{-1}$}
\tablenotetext{b}{Optical counter-parts with no identification}
\tablenotetext{c}{No optical counter-part in either $g^{\prime}$, $r^{\prime}$, or $i^{\prime}$.}
\tablenotetext{d}{Not including clusters are counted under BLAGN, NELG or ALG.}
\end{deluxetable}

\clearpage
\LongTables
\begin{landscape}
\begin{deluxetable}{llllllllllllll}
\tabletypesize{\scriptsize}
\tablewidth{8.6in}
\tablecaption{Hard AGN\tablenotemark{a}\label{catalog}}
\tablehead{
\colhead{Src}&\colhead{ChaMP Name} &\colhead{RA} &\colhead{Dec}
&\colhead{Hard\tablenotemark{b}}&\colhead{log f$_{\rm{x}}$\tablenotemark{c}}
&\colhead{HR} &\colhead{z} &\colhead{log L$_{\rm{x}}$\tablenotemark{d}}
&\colhead{$r^{\prime}$}&\colhead{$g^{\prime}$-$i^{\prime}$} &\colhead{$r^{\prime}$-$i^{\prime}$}
&\colhead{Class}&\colhead{log N$_{\rm{H}}$}\\
\colhead{\#}&&\colhead{(J2000)}&\colhead{(J2000)}&\colhead{counts}&&&&&&&&&\colhead{(cm$^{-2}$)}\\
}
\startdata
1 & CXOMP J001845.7+163346 & 00:18:45.74 & +16:33:46.6 &  89.0$\pm$10.7& --13.51 & --0.40 & 0.624 & 43.70 & 21.41&1.58&0.66&NELG&21.89$^{+0.15}_{-0.18}$\\ 
2 & CXOMP J001837.3+163447 & 00:18:37.38 & +16:34:47.1 &  18.4$\pm$ 5.8& --14.17 & --0.71 & 2.149 & 44.37 & 19.94&0.23&0.15&BLAGN & $<$21.61\\ 
3 & CXOMP J001833.4+163154 & 00:18:33.47 & +16:31:54.4 &  38.0$\pm$ 7.3& --13.90 & --0.65 & 1.643 & 44.36 & 21.11&0.73&0.39&BLAGN & $<$21.58\\ 
4 & CXOMP J001837.4+163757 & 00:18:37.48 & +16:37:57.7 &  19.5$\pm$ 7.3& --14.13 & --0.71 & 1.506 & 44.03 & 20.90&0.45&0.16&BLAGN & $<$22.35\\ 
5 & CXOMP J001828.6+163418 & 00:18:28.68 & +16:34:18.2 &   9.0$\pm$ 4.4& --14.48 & --0.66 & 1.163 & 43.40 & 20.90&0.38&0.04&BLAGN & $<$21.97\\ 
6 & CXOMP J001842.0+163425 & 00:18:42.09 & +16:34:25.0 &   9.3$\pm$ 4.6 & --14.50 & $>$0.04 & 0.550 & 42.59 & 21.23&2.66&0.82&ALG & $<$22.14\\ 
7 & CXOMP J001825.0+163653 & 00:18:25.01 & +16:36:53.0 &  15.6$\pm$ 6.1& --14.25 & --0.49 & 1.198 & 43.67 & 21.56&$<$--2.07&$<$--2.47&BLAGN & $<$22.28\\ 
8 & CXOMP J001850.1+162756 & 00:18:50.14 & +16:27:56.4 &  20.5$\pm$ 5.9& --14.16 & --0.63 & 1.330 & 43.87 & 21.21&0.35&--0.06&BLAGN & $<$22.06\\ 
9 & CXOMP J001859.8+162649 & 00:18:59.83 & +16:26:49.3 &  67.0$\pm$10.0& --13.63 & --0.64 & 1.714 & 44.67 & 19.92&0.31&0.27&BLAGN & $<$22.15\\ 
10 & CXOMP J001854.9+162952 & 00:18:54.92 & +16:29:52.8 &  10.1$\pm$ 4.9& --14.46 & --0.68 & 2.950 & 44.41 & 21.78&0.21&--0.14&BLAGN & $<$22.54\\ 
11 & CXOMP J001810.2+163223 & 00:18:10.24 & +16:32:23.9 &  54.2$\pm$ 8.8& --13.72 & --0.61 & 1.273 & 44.26 & 19.62&0.33&0.01&BLAGN & $<$21.46\\
12 & CXOMP J001801.7+163426 & 00:18:01.72 & +16:34:26.0 &  42.6$\pm$ 9.3& --13.81 & --0.09 & 0.329 & 42.74 & 19.76&1.63&0.34&NELG&22.33$^{+0.20}_{-0.21}$\\
13 & CXOMP J001827.0+162900 & 00:18:27.08 & +16:29:00.1 &  13.1$\pm$ 4.8& --14.33 & --0.55 & 2.828 & 44.49 & 21.27&0.10&--0.16&BLAGN & $<$23.02\\
14 & CXOMP J015327.3--135223 & 01:53:27.30 & --13:52:23.8 & 123.7$\pm$13.3& --13.02 & --0.65 & 0.326 & 43.52 & 19.01&0.88&0.19&BLAGN & $<$21.13\\
15 & CXOMP J015311.1--135104 & 01:53:11.14 & --13:51:04.3 &  64.9$\pm$ 9.7& --13.36 & --0.44 & 0.949 & 44.30 & 21.37&0.32&0.19&BLAGN & $<$21.71\\
16 & CXOMP J015312.3--135723 & 01:53:12.39 & --13:57:23.5 &  27.2$\pm$ 6.7& --13.76 & --0.61 & 1.454 & 44.36 & 20.45&0.57&0.14&BLAGN & $<$21.87\\
17 & CXOMP J015308.0--135801 & 01:53:08.06 & --13:58:01.0 &  13.1$\pm$ 5.0& --14.05 & --0.74 & 1.793 & 44.29 & 19.37&0.23&0.21&BLAGN & $<$21.68\\
18 & CXOMP J015234.7--134735 & 01:52:34.70 & --13:47:35.8 &  23.9$\pm$ 7.9 & --13.77 & $>$0.25 & 0.168 & 42.12 & 17.84&0.93&0.36&BLAGN&21.88$^{+0.74}_{-0.26}$\\
19 & CXOMP J015243.8--135900 & 01:52:43.83 & --13:59:00.7 &  42.7$\pm$ 7.6& --13.57 & --0.72 & 1.674 & 44.71 & 20.88&0.12&0.08&BLAGN & $<$21.14\\
20 & CXOMP J015239.8--135740 & 01:52:39.88 & --13:57:40.3 &  75.9$\pm$ 9.8& --13.31 & --0.31 & 0.868 & 44.25 & 21.47&1.71&0.68&BLAGN&22.31$^{+0.13}_{-0.14}$\\
21 & CXOMP J015234.8--140205 & 01:52:34.87 & --14:02:05.3 &  14.0$\pm$ 5.5& --14.02 & --0.68 & 1.812 & 44.34 & 22.08&0.67&0.45&BLAGN & $<$21.99\\
22 & CXOMP J015234.7--135929 & 01:52:34.76 & --13:59:29.1 &   9.8$\pm$ 4.4& --14.19 & --0.54 & 0.744 & 43.21 & 20.69&0.81&0.43&BLAGN & $<$21.94\\
23 & CXOMP J015235.4--140336 & 01:52:35.45 & --14:03:36.2 &  14.2$\pm$ 6.2& --14.01 & --0.36 & 2.418 & 44.66 & 21.98&0.30&0.07&BLAGN & $<$22.76\\
24 & CXOMP J033752.4--045549 & 03:37:52.44 & --04:55:49.0 &  57.2$\pm$ 9.1& --13.54 & --0.36 & 0.371 & 43.14 & 19.88&1.52&0.56&NELG&21.70$^{+0.27}_{-0.38}$\\
25 & CXOMP J033812.3--050252 & 03:38:12.39 & --05:02:52.3 &  13.7$\pm$ 6.0& --14.16 & --0.63 & 1.464 & 43.97 & 21.76&0.60&0.39&BLAGN & $<$22.05\\
26 & CXOMP J033804.2--050312 & 03:38:04.20 & --05:03:12.5 &  32.4$\pm$ 7.0& --13.79 & --0.62 & 0.587 & 43.37 & 20.62&1.21&0.74&BLAGN & $<$21.17\\
27 & CXOMP J033800.4--050811 & 03:38:00.43 & --05:08:11.4 &  19.9$\pm$ 6.5& --14.01 & --0.55 & 0.434 & 42.82 & 21.09&1.65&0.71&ALG & $<$20.99\\
28 & CXOMP J033723.1--045602 & 03:37:23.12 & --04:56:02.4 &  23.8$\pm$ 7.4& --13.86 & --0.42 & 0.944 & 43.80 & 21.91&1.41&0.72&BLAGN & $<$22.30\\
29 & CXOMP J033731.4--050502 & 03:37:31.44 & --05:05:02.2 &  11.3$\pm$ 4.7 & --14.26 & $>$0.22 & 0.295 & 42.19 & 19.65&1.78&0.64&NELG&23.70$^{+0.47}_{-0.57}$\\
30 & CXOMP J033717.0--050455 & 03:37:17.05 & --05:04:55.4 &  31.8$\pm$ 7.8& --13.78 & --0.46 & 1.859 & 44.60 & 20.79&0.16&0.46&BLAGN & $<$22.34\\
31 & CXOMP J034015.4--352848 & 03:40:15.43 & --35:28:48.6 &  57.6$\pm$ 9.8& --13.45 & --0.79 & 1.737 & 44.86 & 18.80&0.17&0.20&BLAGN & $<$21.33\\
32 & CXOMP J033940.2--353040 & 03:39:40.20 & --35:30:40.3 &  16.7$\pm$ 5.5& --14.04 & --0.74 & 0.454 & 42.84 & 21.08&1.38&0.51&NELG & $<$21.56\\
33 & CXOMP J034019.2--353124 & 03:40:19.21 & --35:31:24.8 &  29.4$\pm$ 9.4& --13.69 & --0.66 & 1.348 & 44.35 & 21.38&0.39&0.06&BLAGN & $<$22.28\\
34 & CXOMP J033942.8--352409 & 03:39:42.89 & --35:24:09.6 &  25.7$\pm$ 6.3& --13.89 & --0.87 & 1.042 & 43.87 & 18.93&0.08&--0.07&BLAGN & $<$20.74\\
35 & CXOMP J033938.2--352351 & 03:39:38.23 & --35:23:51.9 &  16.8$\pm$ 5.3& --14.07 & --0.76 & 0.465 & 42.84 & 21.16&0.98&0.38&BLAGN & $<$21.07\\
36 & CXOMP J033912.1--352811 & 03:39:12.13 & --35:28:11.4 & 158.5$\pm$13.8& --12.78 & --0.68 & 0.463 & 44.12 & 20.11&1.75&0.62&BLAGN & $<$21.54\\
37 & CXOMP J033949.7--352348 & 03:39:49.77 & --35:23:48.8 &  12.2$\pm$ 4.8& --14.21 & --0.40 & 0.522 & 42.82 & 20.41&2.16&0.66&NELG & $<$22.05\\
38 & CXOMP J033934.1--352349 & 03:39:34.19 & --35:23:49.3 &  13.1$\pm$ 4.8 & --14.18 & $>$0.24 & 0.533 & 42.87 & 21.52&2.50&0.95&NELG&23.33$^{+0.78}_{-0.26}$\\
39 & CXOMP J033909.6--352707 & 03:39:09.67 & --35:27:07.3 &  13.3$\pm$ 5.9& --14.14 & --0.77 & 0.525 & 42.89 & 20.61&1.68&0.61&NELG & $<$21.51\\
40 & CXOMP J045452.6--025511 & 04:54:52.68 & --02:55:11.8 &  45.3$\pm$10.5 & --13.57 & $>$0.47 & 0.619 & 43.64 & 21.16&1.85&0.77&NELG&23.01$^{+0.16}_{-0.16}$\\
41 & CXOMP J045407.1--025400 & 04:54:07.10 & --02:54:00.9 &  17.8$\pm$ 5.6& --14.03 & --0.61 & 1.193 & 43.88 & 21.48&0.09&--0.41&BLAGN & $<$21.46\\
42 & CXOMP J045422.1--025124 & 04:54:22.11 & --02:51:24.6 &  49.8$\pm$ 9.1& --13.57 & --0.65 & 0.292 & 42.86 & 19.03&1.73&0.51&BLAGN & $<$21.10\\
43 & CXOMP J045415.4--025429 & 04:54:15.40 & --02:54:29.2 &  10.2$\pm$ 4.6& --14.29 & --0.56 & 0.491 & 42.67 & 21.17&2.22&0.76&ALG & $<$21.68\\
44 & CXOMP J045424.7--025849 & 04:54:24.78 & --02:58:49.1 &  22.9$\pm$ 6.0& --13.94 & --0.43 & 0.724 & 43.43 & 20.99&2.36&1.06&NELG & $<$21.39\\
45 & CXOMP J045422.6--030034 & 04:54:22.62 & --03:00:34.6 &  12.0$\pm$ 4.7& --14.23 & --0.73 & 1.936 & 44.20 & 20.77&0.28&0.34&BLAGN & $<$21.49\\
46 & CXOMP J045419.6--030419 & 04:54:19.66 & --03:04:19.9 & 100.5$\pm$11.5& --13.29 & --0.75 & 0.775 & 44.15 & 18.29&0.34&0.16&BLAGN & $<$20.03\\
47 & CXOMP J045356.3--025837 & 04:53:56.36 & --02:58:37.0 &  95.9$\pm$10.9& --13.29 & --0.51 & 0.129 & 42.35 & 18.67&1.37&0.44&BLAGN & $<$20.90\\
48 & CXOMP J045356.7--030225 & 04:53:56.79 & --03:02:25.6 &  20.2$\pm$ 6.1 & --14.01 & $>$0.50 & 0.423 & 42.80 & 20.05&2.24&0.60&NELG&22.71$^{+0.38}_{-0.33}$\\
49 & CXOMP J054240.8--405626 & 05:42:40.81 & --40:56:26.3 &  46.5$\pm$ 7.9& --13.65 & +0.69 & 0.639 & 43.59 &21.06&3.15&1.06&NELG&23.30$^{+0.13}_{-0.14}$\\
50 & CXOMP J054319.2--405750 & 05:43:19.25 & --40:57:50.5 &  69.4$\pm$ 9.7& --13.49 & --0.71 & 0.337 & 43.09 & 19.33&1.03&0.25&NELG & $<$20.13\\
51 & CXOMP J054328.1--405648 & 05:43:28.12 & --40:56:48.1 &  13.4$\pm$ 6.1& --14.17 & --0.38 & 1.185 & 43.73 & 21.71&1.04&0.31&BLAGN & $<$22.29\\
52 & CXOMP J054330.4--405746 & 05:43:30.47 & --40:57:46.9 &  38.4$\pm$ 8.2& --13.73 & +0.33 & 0.423 & 43.08 &21.70&1.86&0.53&NELG&22.83$^{+0.18}_{-0.19}$\\
53 & CXOMP J054320.6--405619 & 05:43:20.63 & --40:56:19.9 &  29.0$\pm$ 7.1& --13.87 & +0.10 & 0.424 & 42.95 &20.71&1.79&0.55&NELG&22.56$^{+0.15}_{-0.18}$\\
54 & CXOMP J054242.5--405834 & 05:42:42.51 & --40:58:34.0 &  27.4$\pm$ 6.4& --13.86 & --0.73 & 1.442 & 44.25 & 18.85&0.46&0.08&BLAGN & $<$21.40\\
55 & CXOMP J054234.1--405836 & 05:42:34.11 & --40:58:36.7 &  39.6$\pm$ 7.4& --13.69 & --0.45 & 0.948 & 43.97 & 19.35&0.12&--0.14&BLAGN&22.08$^{+0.25}_{-0.34}$\\
56 & CXOMP J054225.9--405846 & 05:42:25.94 & --40:58:46.5 &  33.5$\pm$ 7.0& --13.77 & --0.60 & 1.462 & 44.36 & 20.70&0.67&0.16&BLAGN & $<$21.83\\
57 & CXOMP J054232.8--405627 & 05:42:32.88 & --40:56:27.7 &  18.6$\pm$ 5.6& --14.03 & --0.27 & 1.191 & 43.87 & 22.00&0.75&0.11&BLAGN & $<$22.66\\
58 & CXOMP J054255.0--405956 & 05:42:55.07 & --40:59:56.7 &  32.3$\pm$ 6.9& --13.84 & --0.61 & 2.628 & 44.91 & 20.43&0.22&--0.06&BLAGN & $<$22.25\\
59 & CXOMP J054255.0--405922 & 05:42:55.01 & --40:59:22.7 &  10.4$\pm$ 4.4 & --14.25 & $>$0.28 & 0.644 & 43.00 & 22.25&2.81&0.90&NELG&23.78$^{+0.36}_{-0.86}$\\
60 & CXOMP J054251.4--410205 & 05:42:51.46 & --41:02:05.2 &  23.2$\pm$ 6.0& --13.97 & --0.63 & 0.637 & 43.27 & 19.80&1.22&0.60&ALG & $<$21.70\\
61 & CXOMP J054248.2--410140 & 05:42:48.28 & --41:01:40.3 &  18.5$\pm$ 5.5& --14.07 & --0.62 & 0.634 & 43.16 & 20.81&0.60&0.31&NELG & $<$21.42\\
62 & CXOMP J054259.5--410241 & 05:42:59.55 & --41:02:41.4 &  23.1$\pm$ 6.1& --13.95 & --0.27 & 0.638 & 43.29 & 20.67&1.64&0.67&NELG & $<$22.09\\
63 & CXOMP J054245.6--410607 & 05:42:45.62 & --41:06:07.0 &  17.0$\pm$ 6.3& --14.06 & --0.42 & 0.725 & 43.32 & 22.44&1.74&0.82&BLAGN & $<$21.99\\
64 & CXOMP J054230.5--410405 & 05:42:30.58 & --41:04:05.0 &  13.1$\pm$ 5.6& --14.19 & --0.48 & 1.583 & 44.02 & 21.85&1.86&0.70&BLAGN & $<$22.43\\
65 & CXOMP J062917.6+820534 & 06:29:17.66 & +82:05:34.2 &  77.4$\pm$10.6& --13.38 & --0.58 & 1.101 & 44.44 & 20.76&0.62&0.05&BLAGN & $<$22.06\\
66 & CXOMP J063020.5+815938 & 06:30:20.50 & +81:59:38.0 &  17.6$\pm$ 8.2& --14.02 & --0.81 & 0.742 & 43.38 & 20.31&0.57&0.41&BLAGN & $<$21.18\\
67 & CXOMP J062423.8+820938 & 06:24:23.86 & +82:09:38.2 &  28.9$\pm$ 7.5& --13.84 & --0.63 & 2.585 & 44.89 & 19.60&0.15&0.09&BLAGN & $<$22.36\\
68 & CXOMP J062741.2+820406 & 06:27:41.22 & +82:04:06.9 &  27.3$\pm$ 6.5& --13.87 & --0.64 & 0.259 & 42.44 & 19.47&1.16&0.43&NELG&21.15$^{+0.32}_{-0.78}$\\
69 & CXOMP J062652.8+820441 & 06:26:52.81 & +82:04:41.2 &   9.6$\pm$ 4.4& --14.34 & --0.73 & 2.334 & 44.29 & 20.53&0.66&0.40&BLAGN & $<$21.97\\
70 & CXOMP J062648.6+815932 & 06:26:48.64 & +81:59:32.6 &  10.1$\pm$ 4.6& --14.34 & --0.60 & 2.346 & 44.30 & 22.35&0.06&0.16&BLAGN & $<$22.39\\
71 & CXOMP J074100.8+311853 & 07:41:00.81 & +31:18:53.0 &  13.5$\pm$ 5.2& --13.92 & --0.52 & 0.932 & 43.72 & 21.01&0.75&0.11&BLAGN & $<$22.35\\
72 & CXOMP J074056.1+311208 & 07:40:56.14 & +31:12:08.6 &   9.1$\pm$ 4.3& --14.17 & --0.62 & 0.396 & 42.57 & 21.00&1.43&0.39&BLAGN & $<$21.55\\
73 & CXOMP J084127.2+131411 & 08:41:27.21 & +13:14:11.5 & 164.4$\pm$14.9& --12.91 & --0.75 & 1.545 & 45.28 & 18.93&0.57&0.54&BLAGN & $<$21.12\\
74 & CXOMP J084128.2+131106 & 08:41:28.21 & +13:11:06.8 &  40.9$\pm$ 9.4& --13.48 & --0.66 & 1.866 & 44.91 & 21.04&0.47&0.74&BLAGN & $<$22.65\\
75 & CXOMP J084120.2+130721 & 08:41:20.25 & +13:07:21.8 &  46.9$\pm$ 9.5& --13.44 & --0.61 & 0.303 & 43.03 & 19.13&1.46&0.67&NELG & $<$21.66\\
76 & CXOMP J084054.3+131456 & 08:40:54.30 & +13:14:56.6 &  82.5$\pm$10.2& --13.25 & --0.69 & 0.310 & 43.24 & 19.65&1.50&0.67&BLAGN & $<$21.30\\
77 & CXOMP J084040.8+131726 & 08:40:40.82 & +13:17:26.6 &  24.5$\pm$ 6.3& --13.75 & --0.60 & 2.820 & 45.07 & 20.33&0.25&0.37&BLAGN & $<$22.80\\
78 & CXOMP J084102.9+131716 & 08:41:02.96 & +13:17:16.5 &  33.0$\pm$ 7.2 & --13.63 & $>$0.48 & 0.454 & 43.25 & 21.17&0.60&0.76&NELG&23.29$^{+0.23}_{-0.24}$\\
79 & CXOMP J084052.1+131822 & 08:40:52.12 & +13:18:22.9 &  39.1$\pm$ 7.6& --13.54 & --0.25 & 0.490 & 43.42 & 20.63&2.26&1.03&NELG&22.38$^{+0.19}_{-0.20}$\\
80 & CXOMP J084045.2+131617 & 08:40:45.28 & +13:16:17.4 &  13.3$\pm$ 4.8 & --14.04 & $>$0.11 & 0.573 & 43.09 & 21.72&2.36&1.19&NELG & $<$23.37\\
81 & CXOMP J084102.5+131313 & 08:41:02.58 & +13:13:13.4 &   9.2$\pm$ 4.3& --14.08 & --0.70 & 2.968 & 44.80 & 19.28&0.42&0.52&BLAGN & $<$22.41\\
82 & CXOMP J105623.1--034315 & 10:56:23.12 & --03:43:15.5 &  22.7$\pm$ 9.0& --14.11 & --0.37 & 0.429 & 42.71 & 20.94&1.18&0.30&NELG & $<$22.35\\
83 & CXOMP J105618.1--034240 & 10:56:18.10 & --03:42:40.5 &  43.4$\pm$11.7& --13.89 & --0.55 & 0.181 & 42.07 & 18.93&1.35&0.43&NELG & $<$21.89\\
84 & CXOMP J105626.7--033721 & 10:56:26.77 & --03:37:21.3 &  78.9$\pm$11.5& --13.65 & +0.33 & 0.643 & 43.60 &21.26&2.62&1.04&ALG&23.21$^{+0.15}_{-0.13}$\\
85 & CXOMP J105612.3--033501 & 10:56:12.37 & --03:35:01.5 &  41.6$\pm$15.2& --13.89 & --0.72 & 2.315 & 44.73 & 21.31&0.28&0.22&BLAGN & $<$22.56\\
86 & CXOMP J105643.1--034042 & 10:56:43.11 & --03:40:42.2 & 127.5$\pm$12.4& --13.41 & --0.78 & 2.118 & 45.11 & 18.15&0.27&0.27&BLAGN & $<$21.78\\
87 & CXOMP J105655.1--034322 & 10:56:55.10 & --03:43:22.4 &  19.5$\pm$ 5.8& --14.28 & --0.77 & 4.050 & 44.92 & 22.35&2.22&0.32&BLAGN & $<$22.70\\
88 & CXOMP J105659.4--034716 & 10:56:59.43 & --03:47:15.5 &  59.8$\pm$10.6& --13.77 & --0.59 & 1.784 & 44.57 & 22.27&1.58&0.80&BLAGN&22.46$^{+0.18}_{-0.20}$\\
89 & CXOMP J105646.5--034707 & 10:56:46.38 & --03:47:08.4 &  17.6$\pm$ 8.6& --14.29 & --0.75 & 2.248 & 44.30 & 22.05&0.06&0.09&BLAGN & $<$22.64\\
90 & CXOMP J105655.5--034030 & 10:56:55.59 & --03:40:30.1 &  33.0$\pm$ 7.0& --14.06 & --0.79 & 0.698 & 43.27 & 21.29&0.87&0.58&BLAGN & $<$21.20\\
91 & CXOMP J105646.4--033905 & 10:56:46.47 & --03:39:05.5 &  30.3$\pm$ 6.7& --14.09 & --0.73 & 1.250 & 43.87 & 21.87&0.44&0.20&BLAGN & $<$21.56\\
92 & CXOMP J105700.0--033445 & 10:57:00.06 & --03:34:45.6 &  14.9$\pm$ 5.9& --14.39 & --0.65 & 1.177 & 43.50 & 22.07&0.71&0.24&BLAGN & $<$22.02\\
93 & CXOMP J105650.6--033508 & 10:56:50.63 & --03:35:08.2 &  16.4$\pm$ 5.9& --14.36 & --0.83 & 0.818 & 43.14 & 21.93&0.97&0.67&BLAGN & $<$20.63\\
94 & CXOMP J105641.2--033853 & 10:56:41.29 & --03:38:53.2 &   9.9$\pm$ 4.7& --14.56 & --0.69 & 2.697 & 44.21 & 21.24&0.25&0.12&BLAGN & $<$22.82\\
95 & CXOMP J114215.1+660548 & 11:42:15.19 & +66:05:48.4 &  38.2$\pm$11.3& --14.09 & +0.31 & 0.414 & 42.70 &20.75&2.26&0.67&NELG&22.73$^{+0.18}_{-0.18}$\\
96 & CXOMP J114044.3+660311 & 11:40:44.31 & +66:03:11.8 &  59.3$\pm$ 9.1& --13.94 & --0.68 & 2.310 & 44.67 & 20.56&0.10&0.16&BLAGN & $<$22.06\\
97 & CXOMP J114028.0+660320 & 11:40:28.01 & +66:03:20.1 &  22.6$\pm$ 6.4& --14.34 & --0.62 & 1.957 & 44.10 & 21.64&1.31&0.34&BLAGN & $<$21.97\\
98 & CXOMP J114021.9+660428 & 11:40:21.99 & +66:04:28.8 &  35.0$\pm$ 7.3& --14.17 & --0.69 & 1.413 & 43.92 & 21.27&0.56&--0.12&BLAGN & $<$21.09\\
99 & CXOMP J114024.6+660215 & 11:40:24.60 & +66:02:15.8 &  47.4$\pm$ 8.8& --14.03 & --0.62 & 1.710 & 44.26 & 21.22&0.39&0.54&BLAGN & $<$22.05\\
100 & CXOMP J114022.0+660028 & 11:40:22.00 & +66:00:28.4 &  37.4$\pm$ 9.4& --14.12 & --0.72 & 1.834 & 44.25 & 21.30&0.57&0.20&BLAGN & $<$21.83\\
101 & CXOMP J114018.5+660111 & 11:40:18.55 & +66:01:11.1 &  21.7$\pm$ 7.9& --14.36 & --0.67 & 1.270 & 43.61 & 21.12&0.57&0.36&BLAGN & $<$21.32\\
102 & CXOMP J113950.1+660025 & 11:39:50.17 & +66:00:25.0 &  55.8$\pm$12.9& --13.93 & --0.64 & 0.858 & 43.62 & 21.64&0.51&0.64&BLAGN & $<$21.73\\
103 & CXOMP J114124.3+660921 & 11:41:24.39 & +66:09:21.5 &  61.8$\pm$ 9.2& --13.90 & --0.80 & 1.088 & 43.91 & 19.89&0.28&0.00&BLAGN & $<$20.43\\
104 & CXOMP J114046.4+660913 & 11:40:46.45 & +66:09:13.0 &  16.3$\pm$ 5.3& --14.44 & --0.67 & 1.933 & 43.99 & 21.72&0.03&0.42&BLAGN & $<$22.21\\
105 & CXOMP J114031.1+660858 & 11:40:31.15 & +66:08:58.2 & 198.1$\pm$15.2& --13.41 & --0.65 & 1.269 & 44.56 & 21.02&0.77&0.21&BLAGN & $<$21.56\\
106 & CXOMP J114001.9+660642 & 11:40:01.95 & +66:06:42.3 &  29.2$\pm$ 7.3& --14.24 & --0.64 & 0.481 & 42.70 & 20.25&2.19&0.77&NELG & $<$21.46\\
107 & CXOMP J114036.2+661317 & 11:40:36.22 & +66:13:17.4 &  22.7$\pm$ 7.7& --14.32 & --0.72 & 3.337 & 44.68 & 21.57&0.04&0.10&BLAGN & $<$22.56\\
108 & CXOMP J113944.5+661137 & 11:39:44.58 & +66:11:37.1 &  66.4$\pm$11.7& --13.83 & --0.66 & 2.113 & 44.69 & 20.00&0.98&0.26&BLAGN & $<$21.69\\
109 & CXOMP J113941.2+661319 & 11:39:41.29 & +66:13:19.6 &  34.7$\pm$12.4& --14.11 & --0.56 & 0.495 & 42.86 & 20.76&2.04&0.71&ALG & $<$21.82\\
110 & CXOMP J114022.0+660816 & 11:40:22.00 & +66:08:16.3 &  24.8$\pm$ 7.2& --14.30 & --0.44 & 0.786 & 43.16 & 20.37&3.20&1.42&ALG & $<$21.70\\
111 & CXOMP J134411.0+555353 & 13:44:11.06 & +55:53:53.0 &  12.1$\pm$ 4.8& --14.21 & --0.71 & 0.470 & 42.71 & 20.49&2.30&0.51&NELG & $<$21.76\\
112 & CXOMP J134359.1+555259 & 13:43:59.14 & +55:52:59.6 &  24.1$\pm$ 6.6 & --13.88 & $>$0.34 & 0.593 & 43.28 & 20.67&2.59&0.60&NELG&23.32$^{+0.22}_{-0.24}$\\
113 & CXOMP J134450.6+555531 & 13:44:50.62 & +55:55:31.9 &  25.7$\pm$ 6.3& --13.90 & --0.51 & 1.784 & 44.44 & 20.44&0.35&0.25&BLAGN&22.63$^{+0.19}_{-0.20}$\\
114 & CXOMP J134440.2+555648 & 13:44:40.21 & +55:56:48.4 &  56.0$\pm$ 8.6& --13.55 & --0.78 & 1.156 & 44.32 & 19.05&0.27&0.04&BLAGN & $<$21.14\\
115 & CXOMP J134508.0+555058 & 13:45:08.06 & +55:50:58.6 &  15.9$\pm$ 5.6& --14.11 & --0.65 & 0.608 & 43.08 & 21.08&1.60&0.55&NELG & $<$21.51\\
116 & CXOMP J141656.1+444720 & 14:16:56.18 & +44:47:20.3 &  63.1$\pm$ 9.1& --13.33 & --0.48 & 0.469 & 43.58 & 21.31&2.37&0.74&BLAGN&21.77$^{+0.15}_{-0.16}$\\
117 & CXOMP J141747.4+444544 & 14:17:47.48 & +44:45:44.9 &  18.8$\pm$ 9.4& --13.79 & --0.13 & 2.138 & 44.75 & 19.92&0.34&0.15&BLAGN & $<$22.42\\
118 & CXOMP J141703.7+443851 & 14:17:03.74 & +44:38:51.1 &  26.0$\pm$ 8.8& --13.58 & --0.74 & 1.600 & 44.64 & 19.43&0.56&0.26&BLAGN & $<$21.92\\
119 & CXOMP J141624.9+444045 & 14:16:24.74 & +44:40:45.2 &  35.6$\pm$ 8.4& --13.54 & --0.48 & 1.440 & 44.57 & 20.78&0.82&0.10&BLAGN & $<$21.60\\
120 & CXOMP J141700.7+445606 & 14:17:00.77 & +44:56:06.9 & 575.5$\pm$25.4& --12.30 & --0.77 & 0.114 & 43.22 & 15.93&0.18&0.15&BLAGN & $<$18.99\\
121 & CXOMP J141656.3+445340 & 14:16:56.34 & +44:53:40.2 &  26.5$\pm$ 6.6& --13.67 & +0.25 & 0.440 & 43.18 &19.28&1.85&0.50&NELG&22.80$^{+0.15}_{-0.18}$\\
122 & CXOMP J141655.6+445453 & 14:16:55.62 & +44:54:53.1 &  25.6$\pm$ 6.8& --13.68 & --0.58 & 0.579 & 43.46 & 20.86&2.00&0.53&ALG&21.87$^{+0.20}_{-0.23}$\\
123 & CXOMP J141715.0+445316 & 14:17:15.08 & +44:53:17.1 &  15.1$\pm$ 5.9& --13.91 & --0.73 & 2.449 & 44.77 & 19.62&--0.01&0.03&BLAGN & $<$22.02\\
124 & CXOMP J141637.0+444645 & 14:16:37.08 & +44:46:45.6 &  46.6$\pm$ 7.9& --13.47 & --0.72 & 0.394 & 43.27 & 20.02&0.60&0.23&BLAGN & $<$20.33\\
125 & CXOMP J141624.5+445156 & 14:16:24.52 & +44:51:56.7 &  37.7$\pm$ 7.4& --13.51 & --0.70 & 2.015 & 44.96 & 18.16&0.32&--0.23&BLAGN & $<$21.44\\
126 & CXOMP J141626.6+445240 & 14:16:26.69 & +44:52:40.2 &  26.9$\pm$ 6.6& --13.67 & --0.46 & 0.675 & 43.63 & 20.27&1.55&0.76&BLAGN & $<$21.53\\
127 & CXOMP J141558.7+445009 & 14:15:58.79 & +44:50:09.4 &  29.9$\pm$ 7.9& --13.61 & --0.64 & 1.364 & 44.44 & 19.97&0.19&--0.16&BLAGN & $<$21.17\\
128 & CXOMP J171758.4+671203 & 17:17:58.47 & +67:12:03.1 &  45.2$\pm$ 8.2& --13.69 & --0.62 & 1.800 & 44.66 & 21.05&0.25&0.26&BLAGN & $<$21.84\\
129 & CXOMP J171740.6+671147 & 17:17:40.63 & +67:11:47.2 &  12.3$\pm$ 4.8& --14.29 & --0.57 & 2.318 & 44.33 & 21.97&--0.24&--0.23&BLAGN & $<$22.11\\
130 & CXOMP J171837.7+671351 & 17:18:37.76 & +67:13:51.4 &  20.0$\pm$ 9.3& --14.03 & --0.58 & 1.550 & 44.16 & 20.84&1.03&0.40&BLAGN & $<$22.19\\
131 & CXOMP J171748.3+670544 & 17:17:48.31 & +67:05:44.9 &  27.7$\pm$ 7.6& --13.89 & --0.69 & 0.482 & 43.05 & 19.86&0.35&0.12&BLAGN & $<$21.15\\
132 & CXOMP J171635.5+671626 & 17:16:35.56 & +67:16:26.0 &  21.7$\pm$ 6.1& --13.98 & --0.52 & 0.505 & 43.01 & 20.43&1.97&0.48&ALG & $<$21.70\\
133 & CXOMP J171636.9+670829 & 17:16:36.90 & +67:08:29.7 &  56.2$\pm$ 8.7& --13.59 & +0.51 & 0.795 & 43.88 &22.38&2.75&1.08&NELG&23.20$^{+0.14}_{-0.14}$\\
134 & CXOMP J171700.7+670519 & 17:17:00.70 & +67:05:19.7 &  27.1$\pm$ 7.0& --13.90 & --0.63 & 1.155 & 43.97 & 20.72&0.30&0.01&BLAGN & $<$21.31\\
135 & CXOMP J184514.8+795010 & 18:45:14.81 & +79:50:10.0 &  18.1$\pm$ 7.7& --13.83 & --0.58 & 1.107 & 44.00 & 20.36&0.52&--0.03&BLAGN & $<$21.99\\
136 & CXOMP J205648.1--042937 & 20:56:48.11 & --04:29:37.9 &  34.4$\pm$ 9.0& --13.73 & --0.49 & 0.172 & 42.18 & 18.72&1.73&0.57&NELG & $<$21.74\\
137 & CXOMP J205624.8--042824 & 20:56:24.84 & --04:28:24.4 &  10.1$\pm$ 5.0& --14.23 & --0.58 & 1.511 & 43.93 & 22.13&0.52&0.33&BLAGN & $<$21.68\\
138 & CXOMP J205632.8--042650 & 20:56:32.83 & --04:26:50.4 &  21.0$\pm$ 7.5& --13.95 & --0.52 & 1.511 & 44.22 & 20.18&0.97&0.50&BLAGN & $<$21.95\\
139 & CXOMP J205638.1--043753 & 20:56:38.15 & --04:37:53.2 &  25.6$\pm$ 6.6& --13.85 & --0.54 & 2.970 & 45.03 & 19.70&0.50&0.11&BLAGN & $<$22.66\\
140 & CXOMP J205641.9--043300 & 20:56:41.98 & --04:33:00.7 &  19.4$\pm$ 6.1 & --13.99 & $>$0.48 & 0.467 & 42.92 & 20.91&2.18&0.69&NELG&23.22$^{+0.29}_{-0.31}$\\
141 & CXOMP J205620.5--043059 & 20:56:20.52 & --04:30:59.5 &   8.3$\pm$ 4.1& --14.36 & --0.68 & 2.335 & 44.27 & 21.17&--0.08&0.01&BLAGN & $<$22.44\\
142 & CXOMP J205603.6--043118 & 20:56:03.61 & --04:31:18.0 &  14.8$\pm$ 5.2& --14.08 & --0.66 & 1.010 & 43.65 & 21.12&0.37&0.11&BLAGN & $<$21.25\\
143 & CXOMP J205629.1--043415 & 20:56:29.15 & --04:34:15.8 &   8.7$\pm$ 4.1& --14.35 & --0.74 & 1.031 & 43.40 & 21.42&0.89&0.44&BLAGN & $<$21.32\\
144 & CXOMP J205624.7--043533 & 20:56:24.77 & --04:35:33.8 &  41.4$\pm$ 7.5 & --13.68 & $>$0.73 & 0.261 & 42.64 & 19.48&1.73&0.56&NELG&23.18$^{+0.17}_{-0.19}$\\
145 & CXOMP J205618.6--043429 & 20:56:18.68 & --04:34:29.0 &  35.4$\pm$ 7.1& --13.73 & +0.51 & 0.527 & 43.31 &21.59&2.78&1.01&NELG&23.10$^{+0.14}_{-0.14}$\\
146 & CXOMP J205606.6--043725 & 20:56:06.64 & --04:37:25.1 &  14.5$\pm$ 5.1& --14.11 & --0.76 & 1.188 & 43.79 & 19.93&0.56&0.23&BLAGN & $<$21.37\\
147 & CXOMP J205609.3--043832 & 20:56:09.38 & --04:38:32.2 &  13.9$\pm$ 5.1& --14.12 & --0.24 & 0.396 & 42.62 & 20.91&2.70&0.84&ALG & $<$21.93\\
148 & CXOMP J205608.9--043538 & 20:56:08.93 & --04:35:38.6 &   8.4$\pm$ 4.1 & --14.36 & $>$0.08 & 1.436 & 43.75 & 21.25&0.57&0.18&BLAGN & $<$23.19\\
149 & CXOMP J205605.4--044057 & 20:56:05.47 & --04:40:57.5 &  34.2$\pm$ 7.7& --13.68 & --0.47 & 0.799 & 43.80 & 21.20&0.73&0.52&BLAGN & $<$21.92\\
150 & CXOMP J205603.0--043613 & 20:56:03.03 & --04:36:13.4 &  10.8$\pm$ 4.7& --14.24 & --0.07 & 0.469 & 42.68 & 20.83&2.75&0.94&NELG&22.43$^{+0.22}_{-0.30}$\\
151 & CXOMP J205614.8--044134 & 20:56:14.83 & --04:41:34.8 &  12.8$\pm$ 5.7& --14.13 & --0.62 & 2.476 & 44.56 & 21.95&0.67&0.11&BLAGN & $<$22.71\\
152 & CXOMP J205602.0--043644 & 20:56:02.00 & --04:36:44.8 &  14.3$\pm$ 5.2 & --14.11 & $>$0.42 & 0.466 & 42.80 & 21.74&3.12&0.98&NELG&23.78$^{+0.27}_{-0.34}$\\
153 & CXOMP J213924.9--234221 & 21:39:24.91 & --23:42:21.8 &  33.0$\pm$11.4& --13.68 & --0.43 & 0.401 & 43.08 & 19.73&2.12&0.50&NELG & $<$22.35\\
154 & CXOMP J214041.4--234719 & 21:40:41.47 & --23:47:19.9 & 255.3$\pm$17.9& --12.80 & --0.75 & 0.491 & 44.17 & 18.36&0.11&0.12&BLAGN & $<$20.65\\
155 & CXOMP J214018.0--234920 & 21:40:18.01 & --23:49:20.1 &  17.3$\pm$ 7.7& --13.98 & --0.85 & 1.406 & 44.11 & 19.62&0.28&--0.03&BLAGN & $<$21.71\\
156 & CXOMP J214019.1--234837 & 21:40:19.10 & --23:48:37.7 &  24.9$\pm$ 7.6& --13.83 & +0.01 & 0.387 & 42.88 &21.59&1.79&0.43&NELG&22.83$^{+0.14}_{-0.15}$\\
157 & CXOMP J214018.3--234055 & 21:40:18.32 & --23:40:55.9 &  14.7$\pm$ 5.1& --13.99 & --0.83 & 1.648 & 44.27 & 19.52&0.29&0.15&BLAGN & $<$21.53\\
158 & CXOMP J214010.4--233905 & 21:40:10.49 & --23:39:05.1 &  27.8$\pm$ 6.5 & --13.70 & $>$0.60 & 0.453 & 43.18 & 20.66&2.33&0.58&NELG&23.37$^{+0.19}_{-0.25}$\\
159 & CXOMP J214001.4--234112 & 21:40:01.42 & --23:41:12.8 &  88.5$\pm$10.5& --13.16 & --0.75 & 0.188 & 42.84 & 18.52&1.40&0.40&NELG & $<$20.57\\
160 & CXOMP J224007.1+031813 & 22:40:07.11 & +03:18:13.3 &  84.9$\pm$10.4& --13.15 & --0.72 & 0.528 & 43.89 & 18.96&0.81&0.48&BLAGN & $<$20.54\\
161 & CXOMP J224022.8+032451 & 22:40:22.83 & +03:24:51.3 &  22.4$\pm$ 6.0& --13.67 & --0.71 & 0.679 & 43.64 & 19.92&0.52&0.42&BLAGN & $<$21.07\\
162 & CXOMP J230257.3+084834 & 23:02:57.37 & +08:48:34.8 &  82.0$\pm$10.3& --13.76 & --0.60 & 1.974 & 44.69 & 18.96&0.34&0.33&BLAGN & $<$22.07\\
163 & CXOMP J230252.2+084810 & 23:02:52.22 & +08:48:10.9 &  24.3$\pm$ 6.2& --14.27 & --0.65 & 1.407 & 43.81 & 21.20&0.41&0.18&BLAGN & $<$21.85\\
164 & CXOMP J230252.1+085520 & 23:02:52.17 & +08:55:20.9 &  24.6$\pm$11.6& --14.23 & --0.46 & 3.757 & 44.90 & 20.37&1.70&0.25&BLAGN & $<$23.72\\
165 & CXOMP J230250.9+085311 & 23:02:50.91 & +08:53:11.6 &  23.6$\pm$ 9.0& --14.27 & --0.05 & 0.453 & 42.61 & 20.21&2.11&0.64&ALG&22.27$^{+0.20}_{-0.28}$\\
166 & CXOMP J230300.9+084659 & 23:03:00.98 & +08:46:59.6 & 261.6$\pm$17.3& --13.24 & --0.47 & 0.738 & 44.15 & 21.88&1.34&0.66&BLAGN&21.59$^{+0.14}_{-0.16}$\\
167 & CXOMP J230314.5+084845 & 23:03:14.50 & +08:48:45.0 &  89.4$\pm$12.2 & --13.68 & $>$0.74 & 0.229 & 42.51 & 19.70&2.00&0.67&ALG&23.29$^{+0.18}_{-0.20}$\\
168 & CXOMP J230304.6+084130 & 23:03:04.60 & +08:41:30.4 &  12.7$\pm$ 5.7& --14.51 & --0.61 & 2.592 & 44.22 & 22.06&0.21&0.10&BLAGN & $<$22.45\\
169\tablenotemark{e} & CXOMP J230246.0+084523 & 23:02:46.05 & +08:45:23.8 & 109.3$\pm$11.5& --13.20 & --0.46 & 1.944 & 45.23 & 19.08&0.51&0.55&BLAGN & $\ldots$\\
170 & CXOMP J230221.6+084653 & 23:02:21.66 & +08:46:53.0 &  63.1$\pm$ 9.9& --13.82 & --0.36 & 0.680 & 43.48 & 21.78&1.47&0.66&NELG&21.98$^{+0.17}_{-0.18}$\\
171 & CXOMP J230218.0+084409 & 23:02:18.09 & +08:44:09.4 &  30.8$\pm$ 8.3& --14.13 & --0.65 & 0.988 & 43.57 & 20.38&0.39&0.16&BLAGN & $<$21.37\\
172 & CXOMP J230204.1+084654 & 23:02:04.18 & +08:46:54.8 &  54.8$\pm$13.7& --13.84 & --0.46 & 0.230 & 42.35 & 18.43&1.67&0.62&NELG & $<$21.77\\
173 & CXOMP J230211.1+084654 & 23:02:11.13 & +08:46:57.6 &  27.7$\pm$10.0& --14.16 & --0.18 & 2.570 & 44.57 & 21.35&0.30&0.07&BLAGN&23.34$^{+0.29}_{-0.40}$\\
174 & CXOMP J230254.3+083904 & 23:02:54.38 & +08:39:04.9 & 462.9$\pm$22.9& --12.99 & --0.57 & 0.437 & 43.85 & 19.00&0.91&0.46&BLAGN&21.24$^{+0.11}_{-0.14}$\\
175 & CXOMP J230243.0+083946 & 23:02:43.01 & +08:39:46.7 &  47.3$\pm$ 8.6& --13.98 & --0.53 & 0.438 & 42.86 & 19.26&2.31&0.74&ALG & $<$21.69\\
176 & CXOMP J230240.2+083611 & 23:02:40.21 & +08:36:11.1 & 265.0$\pm$19.0& --13.20 & --0.53 & 1.957 & 45.24 & 20.76&0.67&0.60&BLAGN&22.20$^{+0.12}_{-0.13}$\\
177 & CXOMP J234849.7+010716 & 23:48:49.77 & +01:07:16.2 &  28.3$\pm$10.7& --13.76 & --0.61 & 1.660 & 44.50 & 20.01&0.42&0.23&BLAGN & $<$21.81\\
178 & CXOMP J234812.8+010022 & 23:48:12.89 & +01:00:22.3 &  29.6$\pm$ 6.5& --13.89 & --0.74 & 0.718 & 43.48 & 20.95&0.90&0.50&BLAGN & $<$21.31\\
179 & CXOMP J234826.2+010330 & 23:48:26.26 & +01:03:30.7 &  67.5$\pm$ 9.5& --13.51 & --0.74 & 2.174 & 45.04 & 20.92&0.23&0.21&BLAGN & $<$22.01\\
180 & CXOMP J234823.2+010357 & 23:48:23.26 & +01:03:57.8 &  13.9$\pm$ 5.3& --14.20 & --0.77 & 2.234 & 44.38 & 21.40&--0.11&0.02&BLAGN & $<$22.26\\
181 & CXOMP J234810.5+010552 & 23:48:10.54 & +01:05:51.9 &  27.9$\pm$ 7.5& --13.88 & --0.73 & 1.859 & 44.51 & 19.31&0.32&0.16&BLAGN & $<$22.33\\
182 & CXOMP J234752.5+010306 & 23:47:52.55 & +01:03:06.8 &  15.0$\pm$ 6.5& --14.10 & --0.51 & 2.411 & 44.56 & 20.60&0.36&0.05&BLAGN & $<$22.74\\
183 & CXOMP J234835.3+005832 & 23:48:35.33 & +00:58:32.6 &  84.0$\pm$10.5& --13.52 & --0.72 & 0.946 & 44.14 & 20.87&0.38&0.13&BLAGN & $<$21.37\\
184 & CXOMP J234820.8+010024 & 23:48:20.82 & +01:00:24.2 &  51.0$\pm$ 8.3& --13.80 & --0.75 & 1.210 & 44.12 & 21.31&0.79&0.09&BLAGN & $<$21.07\\
185 & CXOMP J234818.9+005950 & 23:48:18.94 & +00:59:50.1 &  27.3$\pm$ 6.5& --14.08 & --0.77 & 1.937 & 44.35 & 22.14&0.33&0.24&BLAGN & $<$21.55\\
186 & CXOMP J234813.7+005640 & 23:48:13.76 & +00:56:40.0 &  10.0$\pm$ 4.6& --14.51 & --0.11 & 1.042 & 43.25 & 20.37&0.57&0.09&BLAGN&23.16$^{+0.27}_{-0.36}$\\
187 & CXOMP J234811.5+005700 & 23:48:11.54 & +00:57:00.4 &  23.7$\pm$ 6.1& --14.13 & --0.82 & 1.815 & 44.23 & 21.74&0.57&0.42&BLAGN & $<$21.16\\
188 & CXOMP J234820.2+005437 & 23:48:20.21 & +00:54:37.3 &  44.0$\pm$ 8.1& --13.88 & +0.21 & 0.279 & 42.51 &19.33&1.52&0.46&NELG&22.60$^{+0.11}_{-0.13}$\\
\enddata
\tablenotetext{a}{[The complete version of this table is in the electronic edition of the Journal.  The printed edition contains only a sample].}
\tablenotetext{b}{observed frame; 2.5-8.0 keV; background subtracted}
\tablenotetext{c}{galactic absorption-corrected; observed frame; 2.0-8.0 keV; units of erg cm$^{-2}$ s$^{-1}$ }
\tablenotetext{d}{rest frame; 2.0-8.0 keV; units of erg s$^{-1}$}
\tablenotetext{e}{Source fell near the chip gap. No spectral fitting is performed.}


\end{deluxetable}
\clearpage
\end{landscape}

\clearpage

\clearpage

\clearpage

\clearpage

\clearpage

\clearpage

\clearpage

\clearpage

\clearpage

\end{document}